\documentclass[preprint,showpacs,preprintnumbers,amsmath,amssymb]{revtex4}
\usepackage{graphicx}
\usepackage{dcolumn}
\usepackage{bm}

\begin{document}


\title{Phase diagram and symmetry breaking of SU(4) spin-orbital chain in
  a generalized  external field}

\author{Shi-Jian Gu}
\affiliation{Zhejiang Institute of Modern Physics, Zhejiang
University, Hangzhou 310027, P. R. China}

\author{You-Quan Li}
\affiliation{Zhejiang Institute of Modern Physics, Zhejiang
University, Hangzhou 310027, P. R. China}

\author{Huan-Qiang Zhou}
\affiliation{Center for Mathematical Physics, School of Physics
Science, The University of Queensland, 4072, Australia}

\begin{abstract}
The ground state phases of a one-dimensional SU(4) spin-orbital
Hamiltonian in a generalized external field are studied on  the
basis of Bethe-ansatz solution. Introducing three Land\'e $g$
factors for spin, orbital and their products in the SU(4) Zeeman
term, we discuss systematically the various symmetry breaking. The
magnetization versus external field are obtained by solving
Bethe-ansatz equations numerically. The phase diagrams
corresponding to distinct residual symmetries are given by means
of both numerical and analytical methods.
\end{abstract}

\maketitle

\section{Introduction}

There have been much interest in the study of spin models with
orbital degeneracy
\cite{YQLi98,YQLi99,AJoshi99,YTokura00,AMOles00,CItoi00,YLLee00,YYamashita00,SJGu02}
due to experimental progress related to many transition-metal and
rare-earth compounds such as LaMnO$_3$ , CeB$_6$, and perovskite
lattice, as in KCuF$_3$\cite{AMOles00,YTokura00}. Those systems
involve orbital degree of freedom in addition to spin ones. Almost
three decades ago, Kugel and Khomskii\cite{KIKugel73} had pointed
out the possibility of orbital excitations in these systems. As a
model system, it exhibits some fascinating physical features which
is lack without orbital degree of freedom. The isotropic case of
spin system with orbital degeneracy was shown to have an enlarged
SU(4) symmetry \cite{YQLi98}, and one dimensional model is known
to exactly solvable \cite{YQLi99,BSutherland}. Materials related
to spin-orbital systems in one dimension include
quasi-one-dimensional tetrahis-dimethylamino-ethylene
(TDAE)-C$_{60}$\cite{DPArovas95}, artificial quantum dot
arrays\cite{AOnufriev99}, and degenerate chains in
Na$_2$Ti$_2$Sb$_2$O and Na$_2$V$_2$O$_5$ compounds
\cite{EAxtell97,SKPati98}. It is therefore worthwhile to
systematically study the features of one dimensional model.
Theoretical studies\cite{YQLi99} has found the strong interplay of
orbital and spin degrees of freedom in the excitations spectra. It
has been noticed that the presence of orbital may results in
various interesting magnetic properties. Applying a conventional
magnetic field, the spin orbital chain with SU(4) symmetry is
shown to reduce to a model with orbital SU(2)
symmetry\cite{YYamashita00} in the ground state. Recently, we
showed that the magnetization process becomes more complicated if
taking account of the contribution from orbital sector
\cite{SJGu02}. We have explained that the competition between spin
and orbital degree of freedom leads to an orbital
anti-polarization phase. However, the external field we introduced
in ref\cite{SJGu02} is not the most general one for SU(4) systems.

In this paper, we study a SU(4) spin-orbital chain in the presence
of a generalized external field on the basis of its Bethe ansatz
solution. Our paper is organized as follows. In Sec.
\ref{sec:betheansatz}, we introduce the Bethe-ansatz solution and
the Zeeman like term which is going to be added to the original
SU(4) Hamiltonian. In Sec. \ref{sec:GSD}, we give some useful
remarks  on the quantum number configurations for the ground state
in the presence of external field that is characterized by three
parameters. We also simply demonstrate the thermodynamic limit of
the Bethe-ansatz equation and briefly present the dress energy
description of ground state in the presence of external field. In
Sec. \ref{sec:SU3preserved}, we study  the magnetization
properties of a Hamiltonian in the regime with one-parameter
symmetry breaking. In Sec. \ref{sec:SU2preserved}, we study  both
magnetization and the phase diagram in the regimes with
two-parameter symmetry breaking. Various phases and the quantum
phase transitions are obtained by both numerical calculation and
analytical formulation. Concerning to various phases we present
detailed explanation in terms of group theory. Sec.
\ref{sec:discussion} gives a brief summery.

\section{The model and its solution}
\label{sec:betheansatz}

We start from the following Hamiltonian
\begin{equation}
{\cal H}=\sum_{j=1}^N\left[\left(2T_j\cdot
T_{j+1}+\frac{1}{2}\right) \left(2S_j\cdot
S_{j+1}+\frac{1}{2}\right)-1\right]. \label{eq:Hamiltonian}
\end{equation}
where $S_j$ and $T_j$ denote respectively spin and orbital
operators at site $j$, both are generators of SU(2) group
characterizing the spin and orbital degree of freedom of outer
shell electrons in some transitional metal oxides at the
insulating regime.  The coupling constant is set to unit for
simplicity. It has been pointed that the above Hamiltonian
possesses an enlarged SU(4) symmetry\cite{YQLi98} rather than
SU(2)$\times$SU(2) symmetry.

The four states that carry out the fundamental representation of
SU(4) group is denoted by
\begin{eqnarray}
|\underline \uparrow\rangle &=& |1/2, 1/2\rangle,\;\;\;
|\overline\uparrow\rangle = |1/2, -1/2\rangle, \nonumber \\
|\underline \downarrow\rangle &=&|-1/2, 1/2\rangle, \;\;\;
|\overline \downarrow\rangle = |-1/2, -1/2\rangle
\label{eq:Basis}
\end{eqnarray}
These bases are labelled by the eigenvalues of $S^z$ and $T^z$,
{\it i.e.}, $|S^z, T^z\rangle$. As the $su$(4) Lie algebra is of
rank 3, there exists third generator $2S^z T^z$ which possesses
simultaneous eigenvalue  together with $S^z$ and $T^z$.  For
convenience, we denote this new generator by $U^z$ hereafter. In
the terminology of group theory, however, the quadruplet can also
be labelled by the weight vectors which is defined by eigenvalues
of $O_1^z, O_2^z, O_3^z$ that constitute the Cartan subalgebra of
$su(4)$ Lie algebra, Here we adopt the Chevalley basis because the
physical quantities can be conveniently expressed in this basis.

The eigenvalues of $S^z, T^z, U^z$ as well as that of $O_1^z,
O_2^z, O_3^z$  are given in Table I, the relation between these
two basis reads\cite{YQLi99},
\begin{eqnarray}
S^z&=&O_1^z+2O_2^z+O_3^z, \nonumber \\
T^z&=&O_1^z+O_3^z, \nonumber \\
U^z&=&O_1^z-O_3^z.
\end{eqnarray}
\begin{table}
\label{table:dgfgf}
\caption{The eigenvalue of $S^z$, $T^z$, $U^z$
and $z$-component of  $O^z_1, O^z_2, O^z_3$ for the four basis
states (Eq. \ref{eq:Basis}).}
\begin{tabular}{ccccccc}\hline
State  & $\;\;S^z\;\;$ & $\;\;T^z\;\;$ & $\;\;U^z\;\;$
      & $\;\;O_1^z\;\;$ & $\;\;O_2^z\;\;$ & $\;\;O_3^z\;\;$ \\[1mm] \hline
$|\underline \uparrow\rangle$  & 1/2 & 1/2  & 1/2  & 1/2 & 0   & 0
\\[1mm]
$|\overline\uparrow\rangle$    & 1/2 &-1/2  &-1/2  &-1/2 & 1/2 & 0
\\[1mm]
$|\underline\downarrow\rangle$ &-1/2 & 1/2  &-1/2  & 0   &-1/2 &
1/2\\[1mm]
$|\overline \downarrow\rangle$ &-1/2 &-1/2  & 1/2  & 0   & 0
&-1/2\\[1mm]
\hline
\end{tabular}
\end{table}

The present model (\ref{eq:Hamiltonian}) has been solved by
Bethe-ansatz method\cite{BSutherland,YQLi99}, its energy spectrum
is given by
\begin{equation}
E_0(M, M', M'')=-\sum_{a=1}^M\frac{1}{1/4+\lambda_a^2}.
\end{equation}
where the $\lambda$'s are solutions of the following coupled
transcendental equations
\begin{eqnarray}
2\pi I_a &=& N\theta_{-1/2}(\lambda_a)
 +\sum_{a'=1}^M \theta_1(\lambda_a-\lambda_{a'})
   \nonumber\\
&&+\sum_{b=1}^{M'}\theta_{-1/2}(\lambda_a-\mu_b),
 \nonumber \\
2\pi J_b&=& \sum_{a=1}^M\theta_{-1/2}(\mu_b-\lambda_a)
 +\sum_{b'=1}^{M'}\theta_1(\mu_b-\mu_{b'})
  \nonumber \\
& & +\sum_{c=1}^{M''}\theta_{-1/2}(\mu_b-\nu_c),
  \nonumber \\
2\pi
K_c&=&\sum_{b=1}^{M'}\theta_{-1/2}(\nu_c-\mu_b)+\sum_{c'=1}^{M''}\theta_1(\nu_c-\nu_{c'}),
\label{eq:BAE}
\end{eqnarray}
where $\theta_\alpha (x)=-2\tan^{-1}(x/\alpha)$. The $\lambda,
\mu$ and $\nu$ are rapidities related to the three generators of
the Cartan subalgebra of the $su$(4) Lie algebra. The quantum
numbers $\{I_a, J_b, K_c\}$ specify a state in which there are
$N-M$ number of sites in the state $|\underline \uparrow\rangle$,
$M-M'$ in $|\overline\uparrow\rangle$, $M'-M''$ in $|\underline
\downarrow\rangle$, and $M''$  in $|\overline \downarrow\rangle$.
Hence the $z$-component of total spin, orbital and $U^z$ are
obtained as $S_{tot}^z=N/2-M'$, $T_{tot}^z=N/2-M+M'-M''$, and
$U_{tot}^z=N/2-M+M''$.

In present SU(4) model, a three-parameter external field $(h_1,
h_2, h_3 )$ can be introduced to write out a most general
Zeeman-like energy.
\begin{eqnarray}
{\cal H}_{zee}=\sum_m^3 h_m O_m^z. \label{eq:zeemdgkgds}
\end{eqnarray}
For more clear physics implication, we re-choose the parameters to
write the effective magnetization ${\cal M}_z$,
\begin{eqnarray}
{\cal M}^z=g_s S_{tot}^z+g_t T_{tot}^z + g_u U_{tot}^z.
\label{eq:magddgdg}
\end{eqnarray}
where $g_s$, $g_t$, $g_u$ are generalized Land\'e $g$ factors for
$S^z$, $T^z$ and $U^z$ respectively. Eq. (\ref{eq:magddgdg}) can
be expressed in terms of the number of rapidities,
\begin{eqnarray}
{\cal M}^z &=&\frac{N}{2}(g_s+g_t+g_u)-M (g_t+g_u)\nonumber \\
&&-M'(g_s -g_t)-M'' (g_t-g_u), \label{EQ:MAG}
\end{eqnarray}

Because the Zeeman-like term commutes with the SU(4) Hamiltonian
(\ref{eq:Hamiltonian}), the energy spectrum in the presence of
external field are simply related to the energy spectrum in the
absence of external field,
\begin{equation}
E(h, M, M', M'')=E_0(M, M', M'') - h {\cal M}^z,
\label{eq:TEnergy}
\end{equation}
where $E_0(M, M', M'')$ is determined by eqs. (\ref{eq:BAE}).
Obviously, the application of the external field with different
magnitude just brings about various level crossings.

In terms of $O_1, O_2, O_3$, the magnetization (\ref{eq:magddgdg})
becomes
\begin{eqnarray}
{\cal M}^z =(g_s+g_t+g_u) O_1^z +2g_s O_2^z +(g_s+g_t-g_u)
O_3^z.\nonumber\\
\label{eq:mago3}
\end{eqnarray}
which breaks SU(4) symmetry down to various lower symmetries
depending on the distinct regions in the parameter space.

\section{ The Ground state configuration }
\label{sec:GSD}

Based on the Bethe ansatz solution of the model, we first give the
quantum number description of the ground state, which is useful
for numerical method. We also give the dress energy description
for the ground state and propose the conditions to determine
quantum phase transitions, which is useful for analytic study.

It has been known \cite{YQLi98,YQLi99} that the ground state of
the Hamiltonian (\ref{eq:Hamiltonian}) is a SU(4) singlet for the
case of $N=4n$.
The configuration of the quantum number for ground
state
 $\{I_a, J_b, K_c\}$ ($a=1,2,...,3n$; $b=1,2,...,2n$;
$c=1,2,...,n $) are consecutive integers (or half integers)
arranging symmetrically around the zero. In the presence of
magnetic field, however, the Zeeman term brings about level
crossings and the state with $M=3n, M'=2n, M''=n$ is no longer the
ground state. Therefore the numbers $M, M', M''$ for the lowest
energy state are related to the magnitude of the applied external
field.

In order to solve the Bethe ansatz equation numerically, we need
to determine the possible configuration of quantum numbers for
give values $M$, $M'$ and $M''$. The property of Young tableau
requires that  Max$(M)=3N/4$, Max$(M')=N/2$, Max$(M'')=N/4$, and
$N-M \geq M-M' \geq M'-M'' \geq M''$ for a given $N$. Then one is
able to analyze the change of energy level for each state, which
determines the true ground state for a given external field. One
can also calculates the magnetization by Eq. (\ref{EQ:MAG}).

In the thermodynamic limit, the energy (\ref{eq:TEnergy}) is
expressed in terms of densities of the rapidities,
\begin{eqnarray}
 E/N &&= -\frac{h}{2}(g_s+g_t+g_u)
 \nonumber\\
&&  +\int_{-\lambda_0}^{\lambda_0}\sigma(\lambda)[-2\pi
K_{1/2}(\lambda) +(g_t+g_u)h]d\lambda
  \nonumber \\
&&+(g_s-g_t)h\int_{-\mu_0}^{\mu_0 } \omega(\mu)d\mu
  \nonumber\\
&&+(g_t-g_u)h\int_{-\nu_0}^{\nu_0}\tau(\nu)d\nu
 \label{eq:energy}
\end{eqnarray}
These densities satisfy the following coupled integral equations
\begin{eqnarray}
\sigma(\lambda) &=& K_{1/2}(\lambda)-
\int_{-\lambda_0}^{\lambda_0}
K_1(\lambda-\lambda')\sigma(\lambda')d\lambda' \nonumber \\ &&
+ \int_{-\mu_0}^{\mu_0} K_{1/2}(\lambda-\mu)\omega(\mu)d\mu, \nonumber \\
\omega(\mu) &=& \int_{-\lambda_0}^{\lambda_0}
K_{1/2}(\mu-\lambda)\sigma(\lambda)d\lambda \nonumber \\ && -
\int_{-\mu_0}^{\mu_0} K_1(\mu-\mu') \omega(\mu')d\mu'
\nonumber \\ && + \int_{-\nu_0}^{\nu_0}K_{1/2}(\mu-\nu)\tau(\nu)d\nu,\nonumber \\
\tau(\nu) &=& \int_{-\mu_0}^{\mu_0}K_{1/2}(\nu-\mu)\omega(\mu)d\mu
\nonumber \\
&&-\int_{-\nu_0}^{\nu_0}K_1(\nu-\nu')\tau(\nu')d\nu'.
\label{eq:density}
\end{eqnarray}
where $K_n(x)=\pi^{-1} n/ (n^2+x^2)$ and $\lambda_0, \mu_0$ and
$\nu_0$ are determined by
\begin{eqnarray}
\int_{-\lambda_0}^{\lambda_0} \sigma(\lambda) &=& \frac{M}{N}, \nonumber \\
\int_{-\mu_0}^{\mu_0} \omega(\mu) &=& \frac{M'}{N}, \nonumber \\
\int_{-\nu_0}^{\nu_0} \tau(\nu) &=& \frac{M''}{N}.
\end{eqnarray}

It is more convenient to introduce dress energy\cite{HFrahm90}.
The iteration of eq.(\ref{eq:density}) gives rise to
\begin{eqnarray}
\varepsilon(\lambda)&=&-2\pi K_{1/2}(\lambda) +(g_t+g_u)h \nonumber \\
&&- \int_{-\lambda_0}^{\lambda_0} K_1(\lambda-\lambda')
\varepsilon(\lambda')d\lambda' \nonumber \\ &&+ \int_{-\mu_0}^{\mu_0}
K_{1/2}(\lambda-\mu)\zeta(\mu)d\mu. \nonumber \\
\zeta(\mu)&=&(g_s-g_t)h + \int_{-\lambda_0}^{\lambda_0}
K_{1/2}(\mu-\lambda)\varepsilon(\lambda)d\lambda
\nonumber \\
&& -\int_{-\mu_0}^{\mu_0} K_1(\mu-\mu')\zeta(\mu')d\mu' \nonumber \\
&&+\int_{-\nu_0}^{\nu_0} K_{1/2}(\mu-\nu)\xi(\nu)d\nu\nonumber \\
\xi(\nu)&=&(g_t-g_u)h + \int_{-\mu_0}^{\mu_0} K_{1/2}(\nu-\mu)\zeta(\mu)d\mu \nonumber \\
&&-\int_{-\nu_0}^{\nu_0} K_1(\nu-\nu')\xi(\nu')d\nu'.
\label{eq:thermodeq}
\end{eqnarray}
where $\varepsilon, \zeta$, and $\xi$ are the dress energies in
$\lambda$, $\mu$ and $\nu$ sectors respectively. It is worthwhile
to point out that the dress energy is also the thermal potentials
at zero temperature, i.e., $ \exp(\varepsilon/T) =\rho^h/\rho$,
$\exp(\zeta/T)= \sigma^h/\sigma$, and $\exp(\xi/T) =
\omega^h/\omega$. The thermal Bethe-ansatz equation in the
zero-temperature limit turns to eq.(\ref{eq:thermodeq}). In terms
of dress energy, the energy (\ref{eq:energy}) is simplified to
\begin{eqnarray}
E/N=-\frac{h}{2}(g_s+g_t+g_u) +\int^{\lambda_0}_{-\lambda_0}
K_{1/2}(\lambda)\varepsilon(\lambda) d\lambda.
\end{eqnarray}
Apparently, the ground state is a quasi-Dirac sea where the states
of negative dress energy, $\varepsilon(\lambda)<0$,
$\zeta(\mu)<0$, $\xi(\nu)=0$, are fully occupied. The Fermi points
of the three rapidities  are determined by
\begin{eqnarray}
\varepsilon(\lambda_0)=0,\;\; \zeta(\mu_0)=0,\;\; \xi(\nu_0)=0.
\end{eqnarray}
The system will be magnetized if the applied field enhance the
dress energy, because it makes the corresponding Fermi points
decline. The quantum phase transition occurs when any of the Fermi
points shrinks to zero. As a result, the critical values of the
external field are solved by
\begin{eqnarray}
\varepsilon(0)|_{h=h_c}=0,\; \zeta(0)|_{h=h'_c}=0,\;
\xi(0)|_{h=h''_c}=0
\end{eqnarray}
These conditions together with eq.(\ref{eq:thermodeq}) enable one
to calculate those critical values.

\section{Regimes with one-parameter symmetry breaking}
\label{sec:SU3preserved}

The application of external field makes the SU(4) symmetry break
down to various regimes with different residual symmetry. In this
section, we shall discuss the simplest cases of single parameter
hierarchy. There are three special directions in the weight space
of SU(4). If the external field is supplied along those
directions, i.e., either $h_1$, $h_2$ and $h_3$ in eq.
(\ref{eq:zeemdgkgds}) is not vanished, a partial breaking of a
SU(2) to U(1) will take place. Let us consider them respectively.

\subsection{Residual SU(3)$\times$U(1) symmetry}

If $g_s=0, g_t=-g_u>0$, the Zeeman interaction (\ref{eq:mago3})
becomes
\begin{eqnarray}
{\cal M}^z &=& 2g_t O_3^z \nonumber \\
   \,     &=&  g_t M'-2g_t M''.
\end{eqnarray}
The occurrence  of operator $O^z_3$ makes the Hamiltonian
noncommutable with $O^{\pm}_3$. Thus a SU(2) subgroup generated by
$O^z_3$, $O^{+}_3$ $O^{-}_3$ is broken down to U(1). Analyzing the
level crossing from Eq. (\ref{eq:TEnergy}), we shown the
magnetization curve in Fig. \ref{suheis_FIGURE_magn67}.

Because the terms $g_t+g_u$ and $g_s-g_t$ in the first two
equations of eqs. (\ref{eq:thermodeq}) are non-positive, the
external field can not enhance the two dress energies
$\varepsilon(\lambda)$ and $\zeta(\mu)$, the Fermi points in both
$\lambda$ and $\mu$ sectors are fixed. On the contrary, the Zeeman
term has a positive contribution to $\xi(\nu)$, and its two Fermi
points will decline when the external field increases. Although it
is a SU(4) singlet labelled by Young tableau $[n^4]$ in the
absence of external field, the ground state possess a residual
SU(3)$\times$U(1) symmetry in the presence of the aforementioned
one parameter external field at small magnitude, which corresponds
to phase IV labelled by four-row Young tableau. In this regime are
there still three type of rapidities that solve the Bethe-ansatz
equation. The U(1) is generated by $O^z_3$, while the SU(3) is
generated by the following eight operators
\begin{eqnarray}
O^z_1 = \frac{1}{2}(T^z + U^z), &\;& O^z_2 = \frac{1}{2}(S^z
-T^z ),\nonumber\\
O^+_1 = (\frac{1}{2}+S^z )T^+, &\;&       O^+_2 = S^+ T^-,
    \nonumber \\
O^-_1 = (\frac{1}{2}+S^z)T^-,  &\;&      O^-_2 = S^- T^+,
\nonumber \\
O^+_{1+2}=S^+ (\frac{1}{2}+T^z), &\;&   O^-_{1+2}=S^-
(\frac{1}{2}+T^z). \nonumber\\
\end{eqnarray}
There exists a critical field when those two Fermi points shrink
to zero, the rapidity $\nu$ disappears in the Bethe-ansatz
equation. Thus a quantum phase transition occurs at the critical
field which separate two phases, we call phase IV and phase III.

The magnetization process can be clearly illustrated by the
evolution of Young tableau,
\begin{center}
\setlength{\unitlength}{3mm}
\begin{picture}(18,5)(3,0)\linethickness{0.6pt}
\put(-1, 3.35){\vector(0,1){0.7}} \put(-1.2,3.35){\line(1,0){0.4}}
\put(-1, 2.15){\vector(0,1){0.7}} \put(-1.2,2.85){\line(1,0){0.4}}
\put(-1, 1.85){\vector(0,-1){0.7}}\put(-1.2,1.15){\line(1,0){0.4}}
\put(-1, 0.65){\vector(0,-1){0.7}}
\put(-1.2,0.65){\line(1,0){0.4}}

\put(0,1){\line(1,0){1}}
 \put(0,2){\line(1,0){1}}
 \put(0,3){\line(1,0){1}}
 \put(0,0){\line(1,0){1}} \put(3,0){\line(1,0){1}}
  \put(0,4){\line(1,0){1}} \put(3,4){\line(1,0){1}}
\put(3,1){\line(1,0){1}}
 \put(3,2){\line(1,0){1}}
 \put(3,3){\line(1,0){1}}
 \put(1.1,0){\line(1,0){0.1}} \put(1.4,0){\line(1,0){0.1}}  \put(1.7,0){\line(1,0){0.1}}
 \put(2.0,0){\line(1,0){0.1}} \put(2.3,0){\line(1,0){0.1}}  \put(2.6,0){\line(1,0){0.1}}
 \put(2.9,0){\line(1,0){0.1}}
 \put(1.1,1){\line(1,0){0.1}} \put(1.4,1){\line(1,0){0.1}}  \put(1.7,1){\line(1,0){0.1}}
 \put(2.0,1){\line(1,0){0.1}} \put(2.3,1){\line(1,0){0.1}}  \put(2.6,1){\line(1,0){0.1}}
 \put(2.9,1){\line(1,0){0.1}}
 \put(1.1,2){\line(1,0){0.1}} \put(1.4,2){\line(1,0){0.1}}  \put(1.7,2){\line(1,0){0.1}}
 \put(2.0,2){\line(1,0){0.1}} \put(2.3,2){\line(1,0){0.1}}  \put(2.6,2){\line(1,0){0.1}}
 \put(2.9,2){\line(1,0){0.1}}
 \put(1.1,3){\line(1,0){0.1}} \put(1.4,3){\line(1,0){0.1}}  \put(1.7,3){\line(1,0){0.1}}
 \put(2.0,3){\line(1,0){0.1}} \put(2.3,3){\line(1,0){0.1}}  \put(2.6,3){\line(1,0){0.1}}
 \put(2.9,3){\line(1,0){0.1}}
 \put(1.1,4){\line(1,0){0.1}} \put(1.4,4){\line(1,0){0.1}}  \put(1.7,4){\line(1,0){0.1}}
 \put(2.0,4){\line(1,0){0.1}} \put(2.3,4){\line(1,0){0.1}}  \put(2.6,4){\line(1,0){0.1}}
 \put(2.9,4){\line(1,0){0.1}}
  \put(3,1){\line(1,0){1}}
   \put(3,2){\line(1,0){1}}
    \put(3,3){\line(1,0){1}}
 \put(4,0){\line(0,1){4}}
  \put(3,0){\line(0,1){4}}
   \put(1,0){\line(0,1){4}}
   \put(0,0){\line(0,1){4}}

\put(4.5,3.5){\vector(1,0){2}}

\put(7,0){\line(0,1){4}} \put(8,0){\line(0,1){4}}
 \put(9.5,0){\line(0,1){4}} \put(10.5,0){\line(0,1){4}}
    \put(11.5,1){\line(0,1){3}} \put(13,1){\line(0,1){3}}
    \put(14,1){\line(0,1){3}}
\put(7,0){\line(1,0){1}} \put(9.5,0){\line(1,0){1}}
\put(9.5,1){\line(1,0){2}} \put(13,1){\line(1,0){1}}
\put(7,4){\line(1,0){1}} \put(9.5,4){\line(1,0){2}}
\put(13,4){\line(1,0){1}}
  \put(7,1){\line(1,0){1}}
   \put(7,2){\line(1,0){1}}
    \put(7,3){\line(1,0){1}}
    \put(9.5,2){\line(1,0){2}}
    \put(9.5,3){\line(1,0){2}}
     \put(13,2){\line(1,0){1}}
    \put(13,3){\line(1,0){1}}
 \put(8.1,4){\line(1,0){0.1}} \put(8.4,4){\line(1,0){0.1}}  \put(8.7,4){\line(1,0){0.1}}
 \put(9.0,4){\line(1,0){0.1}} \put(9.3,4){\line(1,0){0.1}}
 \put(8.1,3){\line(1,0){0.1}} \put(8.4,3){\line(1,0){0.1}}  \put(8.7,3){\line(1,0){0.1}}
 \put(9.0,3){\line(1,0){0.1}} \put(9.3,3){\line(1,0){0.1}}
 \put(8.1,2){\line(1,0){0.1}} \put(8.4,2){\line(1,0){0.1}}  \put(8.7,2){\line(1,0){0.1}}
 \put(9.0,2){\line(1,0){0.1}} \put(9.3,2){\line(1,0){0.1}}
 \put(8.1,1){\line(1,0){0.1}} \put(8.4,1){\line(1,0){0.1}}  \put(8.7,1){\line(1,0){0.1}}
 \put(9.0,1){\line(1,0){0.1}} \put(9.3,1){\line(1,0){0.1}}
 \put(8.1,0){\line(1,0){0.1}} \put(8.4,0){\line(1,0){0.1}}  \put(8.7,0){\line(1,0){0.1}}
 \put(9.0,0){\line(1,0){0.1}} \put(9.3,0){\line(1,0){0.1}}
 \put(11.6,4){\line(1,0){0.1}} \put(11.9,4){\line(1,0){0.1}}  \put(12.2,4){\line(1,0){0.1}}
 \put(12.5,4){\line(1,0){0.1}} \put(12.8,4){\line(1,0){0.1}}
 \put(11.6,3){\line(1,0){0.1}} \put(11.9,3){\line(1,0){0.1}}  \put(12.2,3){\line(1,0){0.1}}
 \put(12.5,3){\line(1,0){0.1}} \put(12.8,3){\line(1,0){0.1}}
 \put(11.6,2){\line(1,0){0.1}} \put(11.9,2){\line(1,0){0.1}}  \put(12.2,2){\line(1,0){0.1}}
 \put(12.5,2){\line(1,0){0.1}} \put(12.8,2){\line(1,0){0.1}}
 \put(11.6,1){\line(1,0){0.1}} \put(11.9,1){\line(1,0){0.1}}  \put(12.2,1){\line(1,0){0.1}}
 \put(12.5,1){\line(1,0){0.1}} \put(12.8,1){\line(1,0){0.1}}

\put(14.5,3.5){\vector(1,0){2}}
\put(17,1){\line(0,1){3}}\put(18,1){\line(0,1){3}}
\put(20.5,1){\line(0,1){3}} \put(21.5,1){\line(0,1){3}}
\put(17,1){\line(1,0){1}} \put(20.5,1){\line(1,0){1}}
\put(17,4){\line(1,0){1}} \put(20.5,4){\line(1,0){1}}
 \put(17,2){\line(1,0){1}}
 \put(17,3){\line(1,0){1}}
 \put(20.5,2){\line(1,0){1}}
 \put(20.5,3){\line(1,0){1}}
 \put(18.1,4){\line(1,0){0.1}} \put(18.4,4){\line(1,0){0.1}}  \put(18.7,4){\line(1,0){0.1}}
 \put(19.0,4){\line(1,0){0.1}} \put(19.3,4){\line(1,0){0.1}}  \put(19.6,4){\line(1,0){0.1}}
 \put(19.9,4){\line(1,0){0.1}} \put(20.2,4){\line(1,0){0.1}}
 \put(18.1,3){\line(1,0){0.1}} \put(18.4,3){\line(1,0){0.1}}  \put(18.7,3){\line(1,0){0.1}}
 \put(19.0,3){\line(1,0){0.1}} \put(19.3,3){\line(1,0){0.1}}  \put(19.6,3){\line(1,0){0.1}}
 \put(19.9,3){\line(1,0){0.1}} \put(20.2,3){\line(1,0){0.1}}
 \put(18.1,2){\line(1,0){0.1}} \put(18.4,2){\line(1,0){0.1}}  \put(18.7,2){\line(1,0){0.1}}
 \put(19.0,2){\line(1,0){0.1}} \put(19.3,2){\line(1,0){0.1}}  \put(19.6,2){\line(1,0){0.1}}
 \put(19.9,2){\line(1,0){0.1}} \put(20.2,2){\line(1,0){0.1}}
 \put(18.1,1){\line(1,0){0.1}} \put(18.4,1){\line(1,0){0.1}}  \put(18.7,1){\line(1,0){0.1}}
 \put(19.0,1){\line(1,0){0.1}} \put(19.3,1){\line(1,0){0.1}}  \put(19.6,1){\line(1,0){0.1}}
 \put(19.9,1){\line(1,0){0.1}} \put(20.2,1){\line(1,0){0.1}}
\end{picture}
\end{center}

which shows the evolution of Young tableau in magnetization
process, i.e., from SU(4) singlet to SU(3)$\times$U(1) states and
then to a SU(3) singlet. Physically, we have ${\cal M}^z /N=0$ at
zero external field due to there are a quart of the total sites
being respectively in the states $|\underline \uparrow\rangle$,
$|\overline\uparrow\rangle$,  $|\underline\downarrow\rangle$ and
$|\overline \downarrow\rangle$. Turning on the external field
leads to the spin-orbital flipping, $|\overline
\downarrow\rangle\rightarrow |\underline \uparrow\rangle$,
$|\overline \downarrow\rangle \rightarrow
|\overline\uparrow\rangle$ and $|\overline \downarrow\rangle
\rightarrow | \underline\downarrow\rangle$, which result in
non-vanishing magnetization. The SU(3)$\times$U(1) symmetry makes
the above three flipping processes favor to occur simultaneously.
When the external field succeeds a critical value, it goes into
phase III where all the states $|\overline \downarrow\rangle$ have
been flipped over. In this phase the $z$-component of total spin
and total orbital keep positive constant $S^z/N=T^z/N=1/6$ while
that of $U^z$ keeps a negative constant $U^z/N=-1/6$.
Consequently, the magnetization reaches a saturate value ${\cal
M}^z /N=2/3$ and the ground state becomes the SU(3) singlet
regardless of the magnitude of the external field in phase III.

\subsection{ Residual SU(2)$\times$SU(2) symmetry}

Applying the external field along the direction of the second
simple root of  $su(4)$ Lie algebra, we will have a symmetry
breaking from SU(4) to SU(2)$\times$U(1)$\times$SU(2) for ground
state. This is realized by the choice of Land\'e g factors $g_u=0,
g_s=-g_t$, which makes the magnetization to be
\begin{eqnarray}
{\cal M}^z &=& 2g_s O^z_2\nonumber\\
 \,        &=&g_s M - 2g_s M'+g_s M''.
\end{eqnarray}
Those two SU(2) are generated respectively by
\begin{eqnarray}
&&\left\{\frac{1}{2}(T^z +U^z), \, (\frac{1}{2} +S^z)T^\pm
\right\},
  \nonumber \\
&& \left\{ \frac{1}{2}(T^z -U^z), \, (\frac{1}{2}-S^z)T^\pm
\right\}.
\end{eqnarray}
As $g_s-g_t$ in the second equation of (\ref{eq:thermodeq}) is
positive but both $g_t+g_u$ and $g_t-g_u$ in the first and third
equation are negative, the external field makes the Fermi points
in $\mu$ sector to shrink. The critical value of the external
field when quantum phase transition occurs is determined by
$\zeta(0)|_{h_c}=0$. This critical point separates two different
phases, we call phase IV and phase II.

The magnetization curve  is shown in Fig.
\ref{suheis_FIGURE_magn67}. It can also be illustrated by the
evolution of the Young tableau,
\begin{center}
\setlength{\unitlength}{3mm}
\begin{picture}(18,5)(3,0)\linethickness{0.6pt}
\put(-1, 3.35){\vector(0,1){0.7}} \put(-1.2,3.35){\line(1,0){0.4}}
\put(-1, 2.15){\vector(0,1){0.7}} \put(-1.2,2.85){\line(1,0){0.4}}
\put(-1, 1.85){\vector(0,-1){0.7}}\put(-1.2,1.15){\line(1,0){0.4}}
\put(-1, 0.65){\vector(0,-1){0.7}}
\put(-1.2,0.65){\line(1,0){0.4}}

\put(0,1){\line(1,0){1}}
 \put(0,2){\line(1,0){1}}
 \put(0,3){\line(1,0){1}}
 \put(0,0){\line(1,0){1}} \put(3,0){\line(1,0){1}}
  \put(0,4){\line(1,0){1}} \put(3,4){\line(1,0){1}}
\put(3,1){\line(1,0){1}}
 \put(3,2){\line(1,0){1}}
 \put(3,3){\line(1,0){1}}
 \put(1.1,0){\line(1,0){0.1}} \put(1.4,0){\line(1,0){0.1}}  \put(1.7,0){\line(1,0){0.1}}
 \put(2.0,0){\line(1,0){0.1}} \put(2.3,0){\line(1,0){0.1}}  \put(2.6,0){\line(1,0){0.1}}
 \put(2.9,0){\line(1,0){0.1}}
 \put(1.1,1){\line(1,0){0.1}} \put(1.4,1){\line(1,0){0.1}}  \put(1.7,1){\line(1,0){0.1}}
 \put(2.0,1){\line(1,0){0.1}} \put(2.3,1){\line(1,0){0.1}}  \put(2.6,1){\line(1,0){0.1}}
 \put(2.9,1){\line(1,0){0.1}}
 \put(1.1,2){\line(1,0){0.1}} \put(1.4,2){\line(1,0){0.1}}  \put(1.7,2){\line(1,0){0.1}}
 \put(2.0,2){\line(1,0){0.1}} \put(2.3,2){\line(1,0){0.1}}  \put(2.6,2){\line(1,0){0.1}}
 \put(2.9,2){\line(1,0){0.1}}
 \put(1.1,3){\line(1,0){0.1}} \put(1.4,3){\line(1,0){0.1}}  \put(1.7,3){\line(1,0){0.1}}
 \put(2.0,3){\line(1,0){0.1}} \put(2.3,3){\line(1,0){0.1}}  \put(2.6,3){\line(1,0){0.1}}
 \put(2.9,3){\line(1,0){0.1}}
 \put(1.1,4){\line(1,0){0.1}} \put(1.4,4){\line(1,0){0.1}}  \put(1.7,4){\line(1,0){0.1}}
 \put(2.0,4){\line(1,0){0.1}} \put(2.3,4){\line(1,0){0.1}}  \put(2.6,4){\line(1,0){0.1}}
 \put(2.9,4){\line(1,0){0.1}}
  \put(3,1){\line(1,0){1}}
   \put(3,2){\line(1,0){1}}
    \put(3,3){\line(1,0){1}}
 \put(4,0){\line(0,1){4}}
  \put(3,0){\line(0,1){4}}
   \put(1,0){\line(0,1){4}}
   \put(0,0){\line(0,1){4}}

\put(4.5,3.5){\vector(1,0){2}}

\put(7,0){\line(0,1){4}} \put(8,0){\line(0,1){4}}
 \put(9.5,0){\line(0,1){4}} \put(10.5,0){\line(0,1){4}}
    \put(11.5,2){\line(0,1){2}} \put(13,2){\line(0,1){2}}
    \put(14,2){\line(0,1){2}}
\put(7,0){\line(1,0){1}} \put(9.5,0){\line(1,0){1}}
\put(9.5,1){\line(1,0){1}} \put(7,4){\line(1,0){1}}
\put(9.5,4){\line(1,0){2}} \put(13,4){\line(1,0){1}}
  \put(7,1){\line(1,0){1}}
   \put(7,2){\line(1,0){1}}
    \put(7,3){\line(1,0){1}}
    \put(9.5,2){\line(1,0){2}}
    \put(9.5,3){\line(1,0){2}}
     \put(13,2){\line(1,0){1}}
    \put(13,3){\line(1,0){1}}
 \put(8.1,4){\line(1,0){0.1}} \put(8.4,4){\line(1,0){0.1}}  \put(8.7,4){\line(1,0){0.1}}
 \put(9.0,4){\line(1,0){0.1}} \put(9.3,4){\line(1,0){0.1}}
 \put(8.1,3){\line(1,0){0.1}} \put(8.4,3){\line(1,0){0.1}}  \put(8.7,3){\line(1,0){0.1}}
 \put(9.0,3){\line(1,0){0.1}} \put(9.3,3){\line(1,0){0.1}}
 \put(8.1,2){\line(1,0){0.1}} \put(8.4,2){\line(1,0){0.1}}  \put(8.7,2){\line(1,0){0.1}}
 \put(9.0,2){\line(1,0){0.1}} \put(9.3,2){\line(1,0){0.1}}
 \put(8.1,1){\line(1,0){0.1}} \put(8.4,1){\line(1,0){0.1}}  \put(8.7,1){\line(1,0){0.1}}
 \put(9.0,1){\line(1,0){0.1}} \put(9.3,1){\line(1,0){0.1}}
 \put(8.1,0){\line(1,0){0.1}} \put(8.4,0){\line(1,0){0.1}}  \put(8.7,0){\line(1,0){0.1}}
 \put(9.0,0){\line(1,0){0.1}} \put(9.3,0){\line(1,0){0.1}}
 \put(11.6,4){\line(1,0){0.1}} \put(11.9,4){\line(1,0){0.1}}  \put(12.2,4){\line(1,0){0.1}}
 \put(12.5,4){\line(1,0){0.1}} \put(12.8,4){\line(1,0){0.1}}
 \put(11.6,3){\line(1,0){0.1}} \put(11.9,3){\line(1,0){0.1}}  \put(12.2,3){\line(1,0){0.1}}
 \put(12.5,3){\line(1,0){0.1}} \put(12.8,3){\line(1,0){0.1}}
 \put(11.6,2){\line(1,0){0.1}} \put(11.9,2){\line(1,0){0.1}}  \put(12.2,2){\line(1,0){0.1}}
 \put(12.5,2){\line(1,0){0.1}} \put(12.8,2){\line(1,0){0.1}}

\put(14.5,3.5){\vector(1,0){2}}
\put(17,2){\line(0,1){2}}\put(18,2){\line(0,1){2}}
\put(20.5,2){\line(0,1){2}} \put(21.5,2){\line(0,1){2}}
\put(17,4){\line(1,0){1}} \put(20.5,4){\line(1,0){1}}
 \put(17,2){\line(1,0){1}}
 \put(17,3){\line(1,0){1}}
 \put(20.5,2){\line(1,0){1}}
 \put(20.5,3){\line(1,0){1}}
 \put(18.1,4){\line(1,0){0.1}} \put(18.4,4){\line(1,0){0.1}}  \put(18.7,4){\line(1,0){0.1}}
 \put(19.0,4){\line(1,0){0.1}} \put(19.3,4){\line(1,0){0.1}}  \put(19.6,4){\line(1,0){0.1}}
 \put(19.9,4){\line(1,0){0.1}} \put(20.2,4){\line(1,0){0.1}}
 \put(18.1,3){\line(1,0){0.1}} \put(18.4,3){\line(1,0){0.1}}  \put(18.7,3){\line(1,0){0.1}}
 \put(19.0,3){\line(1,0){0.1}} \put(19.3,3){\line(1,0){0.1}}  \put(19.6,3){\line(1,0){0.1}}
 \put(19.9,3){\line(1,0){0.1}} \put(20.2,3){\line(1,0){0.1}}
 \put(18.1,2){\line(1,0){0.1}} \put(18.4,2){\line(1,0){0.1}}  \put(18.7,2){\line(1,0){0.1}}
 \put(19.0,2){\line(1,0){0.1}} \put(19.3,2){\line(1,0){0.1}}  \put(19.6,2){\line(1,0){0.1}}
 \put(19.9,2){\line(1,0){0.1}} \put(20.2,2){\line(1,0){0.1}}
\end{picture}
\end{center}
The spin-orbital flipping process caused by the applied external
field has several characteristics. During the flipping process,
$|\underline\downarrow\rangle$ and $|\overline\downarrow\rangle$
flips simultaneously into $|\overline\uparrow\rangle$ and
$|\underline\uparrow\rangle$ pairs, which makes four-row Young
tableau reduce to the two-row Young tableau when across the
critical field. Apparently, the eigenvalues of both $T^z$ and
$U^z$ do not change during the magnetization process. Only $S^z$
contributes to the magnetization ${\cal M}^z $. The total spin is
completely polarized (i.e., ${\cal M}^z$ is saturated) once the
phase IV transits to phase II. The phase that Yamashita {\it et al
}\cite{YYamashita00} discussed agrees with this special case.

\subsection{Residual U(1)$\times$SU(3) symmetry}

\begin{figure*}
\includegraphics[width=15cm]{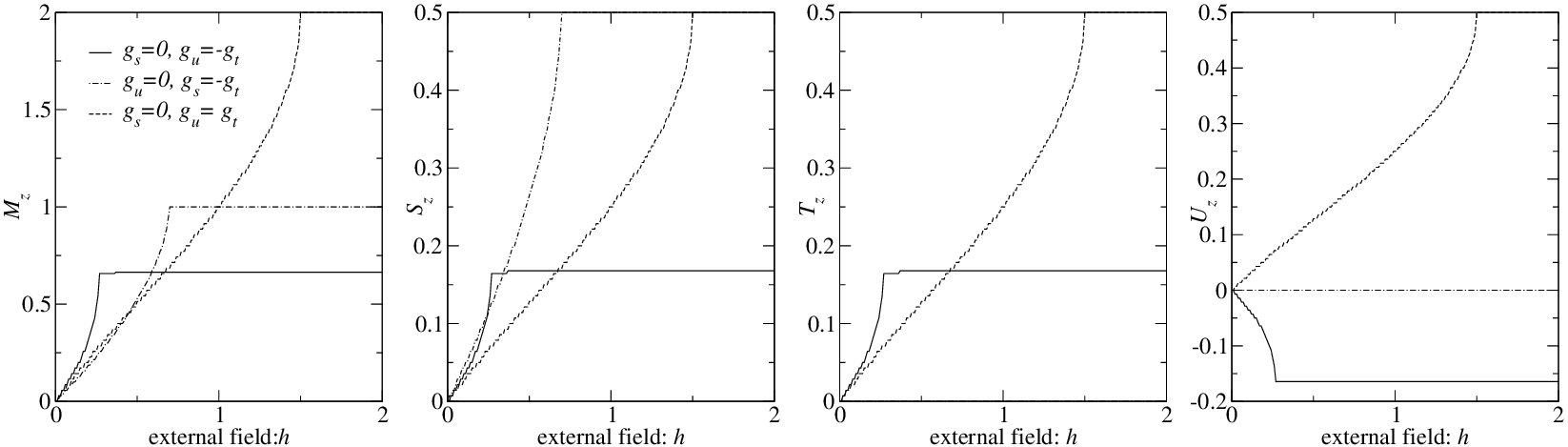}
\caption{\label{suheis_FIGURE_magn67} The magnetization ${\cal
M}^z , S_z, T_z, U_z$ of the system in with (1) $g_s=0, g_t=-g_u$;
(2) $g_u=0, g_s=-g_t$, and (3) $g_s=0$, $g_t=g_u$.}
\end{figure*}

If $g_s=0$ and $g_t=g_u$, the magnetization becomes
\begin{eqnarray}
{\cal M}^z &=& 2g_t O^z_1\nonumber\\
           &=& g_t (N-2M+M').
\end{eqnarray}
which implies that the external field was applied along the first
simple root of $su(4)$ Lie algebra. This gives rise to a symmetry
breaking down to U(1)$\times$SU(3) for ground state. For the sake
of saving space, we omitted the operators that generate those
symmetry. Because this parameter choice implies that $g_t+g_u$ in
$\lambda$ sector positive but the $g$ factor terms in both $\mu$
and $\nu$ sectors are non-positive, the quantum phase transition
is only related to $\lambda$ sector. The critical value is
determined by $\varepsilon(0)|_{h_c}=0$. This critical point
separates the system into two phases, phase IV and phase I. The
magnetization process is shown in Fig.\ref{suheis_FIGURE_magn67}.

It is helpful to illustrated this process  by the evolution of
Young tableau,
\begin{center}
\setlength{\unitlength}{3mm}
\begin{picture}(18,5)(3,0)\linethickness{0.6pt}
\put(-1, 3.35){\vector(0,1){0.7}} \put(-1.2,3.35){\line(1,0){0.4}}
\put(-1, 2.15){\vector(0,1){0.7}} \put(-1.2,2.85){\line(1,0){0.4}}
\put(-1, 1.85){\vector(0,-1){0.7}}\put(-1.2,1.15){\line(1,0){0.4}}
\put(-1, 0.65){\vector(0,-1){0.7}}
\put(-1.2,0.65){\line(1,0){0.4}}

\put(0,1){\line(1,0){1}}
 \put(0,2){\line(1,0){1}}
 \put(0,3){\line(1,0){1}}
 \put(0,0){\line(1,0){1}}  \put(3,0){\line(1,0){1}}
  \put(0,4){\line(1,0){1}} \put(3,4){\line(1,0){1}}
\put(3,1){\line(1,0){1}}
 \put(3,2){\line(1,0){1}}
 \put(3,3){\line(1,0){1}}
 \put(1.1,0){\line(1,0){0.1}} \put(1.4,0){\line(1,0){0.1}}  \put(1.7,0){\line(1,0){0.1}}
 \put(2.0,0){\line(1,0){0.1}} \put(2.3,0){\line(1,0){0.1}}  \put(2.6,0){\line(1,0){0.1}}
 \put(2.9,1){\line(1,0){0.1}}
 \put(1.1,1){\line(1,0){0.1}} \put(1.4,1){\line(1,0){0.1}}  \put(1.7,1){\line(1,0){0.1}}
 \put(2.0,1){\line(1,0){0.1}} \put(2.3,1){\line(1,0){0.1}}  \put(2.6,1){\line(1,0){0.1}}
 \put(2.9,1){\line(1,0){0.1}}
 \put(1.1,2){\line(1,0){0.1}} \put(1.4,2){\line(1,0){0.1}}  \put(1.7,2){\line(1,0){0.1}}
 \put(2.0,2){\line(1,0){0.1}} \put(2.3,2){\line(1,0){0.1}}  \put(2.6,2){\line(1,0){0.1}}
 \put(2.9,2){\line(1,0){0.1}}
 \put(1.1,3){\line(1,0){0.1}} \put(1.4,3){\line(1,0){0.1}}  \put(1.7,3){\line(1,0){0.1}}
 \put(2.0,3){\line(1,0){0.1}} \put(2.3,3){\line(1,0){0.1}}  \put(2.6,3){\line(1,0){0.1}}
 \put(2.9,3){\line(1,0){0.1}}
 \put(1.1,4){\line(1,0){0.1}} \put(1.4,4){\line(1,0){0.1}}  \put(1.7,4){\line(1,0){0.1}}
 \put(2.0,4){\line(1,0){0.1}} \put(2.3,4){\line(1,0){0.1}}  \put(2.6,4){\line(1,0){0.1}}
 \put(2.9,4){\line(1,0){0.1}}
  \put(3,1){\line(1,0){1}}
   \put(3,2){\line(1,0){1}}
    \put(3,3){\line(1,0){1}}
 \put(4,0){\line(0,1){4}}
  \put(3,0){\line(0,1){4}}
   \put(1,0){\line(0,1){4}}
   \put(0,0){\line(0,1){4}}

\put(4.5,3.5){\vector(1,0){2}}

\put(7,0){\line(0,1){4}} \put(8,0){\line(0,1){4}}
 \put(9.5,0){\line(0,1){4}} \put(10.5,0){\line(0,1){4}}
    \put(11.5,3){\line(0,1){1}} \put(13,3){\line(0,1){1}}
    \put(14,3){\line(0,1){1}}
\put(7,0){\line(1,0){1}} \put(9.5,0){\line(1,0){1}}
\put(9.5,3){\line(1,0){2}} \put(13,3){\line(1,0){1}}
\put(7,4){\line(1,0){1}} \put(9.5,4){\line(1,0){2}}
\put(13,4){\line(1,0){1}}
  \put(7,1){\line(1,0){1}}
   \put(7,2){\line(1,0){1}}
    \put(7,3){\line(1,0){1}}
    \put(9.5,2){\line(1,0){1}}
    \put(9.5,1){\line(1,0){1}}
 \put(8.1,4){\line(1,0){0.1}} \put(8.4,4){\line(1,0){0.1}}  \put(8.7,4){\line(1,0){0.1}}
 \put(9.0,4){\line(1,0){0.1}} \put(9.3,4){\line(1,0){0.1}}
 \put(8.1,3){\line(1,0){0.1}} \put(8.4,3){\line(1,0){0.1}}  \put(8.7,3){\line(1,0){0.1}}
 \put(9.0,3){\line(1,0){0.1}} \put(9.3,3){\line(1,0){0.1}}
 \put(8.1,2){\line(1,0){0.1}} \put(8.4,2){\line(1,0){0.1}}  \put(8.7,2){\line(1,0){0.1}}
 \put(9.0,2){\line(1,0){0.1}} \put(9.3,2){\line(1,0){0.1}}
 \put(8.1,1){\line(1,0){0.1}} \put(8.4,1){\line(1,0){0.1}}  \put(8.7,1){\line(1,0){0.1}}
 \put(9.0,1){\line(1,0){0.1}} \put(9.3,1){\line(1,0){0.1}}
 \put(8.1,0){\line(1,0){0.1}} \put(8.4,0){\line(1,0){0.1}}  \put(8.7,0){\line(1,0){0.1}}
 \put(9.0,0){\line(1,0){0.1}} \put(9.3,0){\line(1,0){0.1}}

 \put(11.6,4){\line(1,0){0.1}} \put(11.9,4){\line(1,0){0.1}}  \put(12.2,4){\line(1,0){0.1}}
 \put(12.5,4){\line(1,0){0.1}} \put(12.8,4){\line(1,0){0.1}}
 \put(11.6,3){\line(1,0){0.1}} \put(11.9,3){\line(1,0){0.1}}  \put(12.2,3){\line(1,0){0.1}}
 \put(12.5,3){\line(1,0){0.1}} \put(12.8,3){\line(1,0){0.1}}

\put(14.5,3.5){\vector(1,0){2}}
\put(17,3){\line(0,1){1}}\put(18,3){\line(0,1){1}}
\put(20.5,3){\line(0,1){1}} \put(21.5,3){\line(0,1){1}}
\put(17,3){\line(1,0){1}} \put(20.5,3){\line(1,0){1}}
\put(17,4){\line(1,0){1}} \put(20.5,4){\line(1,0){1}}

 \put(18.1,4){\line(1,0){0.1}} \put(18.4,4){\line(1,0){0.1}}  \put(18.7,4){\line(1,0){0.1}}
 \put(19.0,4){\line(1,0){0.1}} \put(19.3,4){\line(1,0){0.1}}  \put(19.6,4){\line(1,0){0.1}}
 \put(19.9,4){\line(1,0){0.1}} \put(20.2,4){\line(1,0){0.1}}
 \put(18.1,3){\line(1,0){0.1}} \put(18.4,3){\line(1,0){0.1}}  \put(18.7,3){\line(1,0){0.1}}
 \put(19.0,3){\line(1,0){0.1}} \put(19.3,3){\line(1,0){0.1}}  \put(19.6,3){\line(1,0){0.1}}
 \put(19.9,3){\line(1,0){0.1}} \put(20.2,3){\line(1,0){0.1}}
\end{picture}
\end{center}
From Fig.\ref{suheis_FIGURE_magn67}, we see that the spin, orbital
as well as $U^z$ are polarized simultaneously versus external
field. The above Young tableau indicates that the states
$|\overline\uparrow\rangle, |\underline \downarrow\rangle,
|\overline \downarrow\rangle$ change to the state $|\underline
\uparrow\rangle$ simultaneously because they carries out a SU(3)
representation and the system possess SU(3) symmetry. After the
system is fully polarized, the magnetization reach the maximum
value ${\cal M}^z/N =1/2$. Then the residual symmetry of the
ground state is only of U(1).

\section{Regimes with two-parameter symmetry breaking }
\label{sec:SU2preserved}

In the previous section, we discussed the simplest case where
merely one SU(2) subgroup symmetry is broken. In  the following,
we will consider the regimes with two-parameter symmetry breaking,
which involves more SU(2) subgroup.

\subsection{Residual U(1)$\times$ U(1) $\times$ SU(2) symmetry}

Under the restriction $g_u=g_s+g_t$, the magnetization
(\ref{EQ:MAG}) becomes,
\begin{eqnarray}
{\cal M}^z =2(g_s+g_t)O_1^z+2g_s O_2^z
\end{eqnarray}
which indicates that the residual symmetry of ground state is
U(1)$\times$ U(1) $\times$ SU(2), in which the SU(2) is generated
by $O_3^z=(T^z-U^z)/2$ and $O_3^\pm =T^\pm (1/2-S^z)$.

\begin{figure*}
\includegraphics[width=15cm]{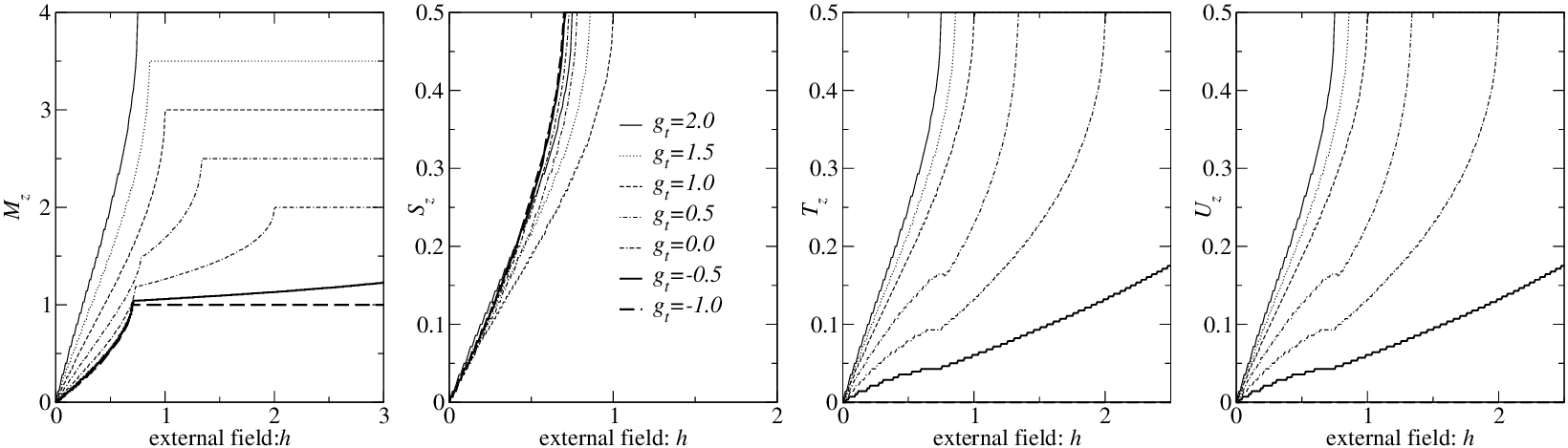}
\caption{\label{suheis_FIGURE_magn3} The magnetization ${\cal M}^z
, S_z, T_z, U_z$ of the system with $g_s=2, g_u=g_s+g_t$, and
$g_t=2.0, 1.5, 1.0, 0.5, 0.0, -0.5, -1.0$.\\\\}
\end{figure*}

\begin{figure}
\includegraphics[width=7cm]{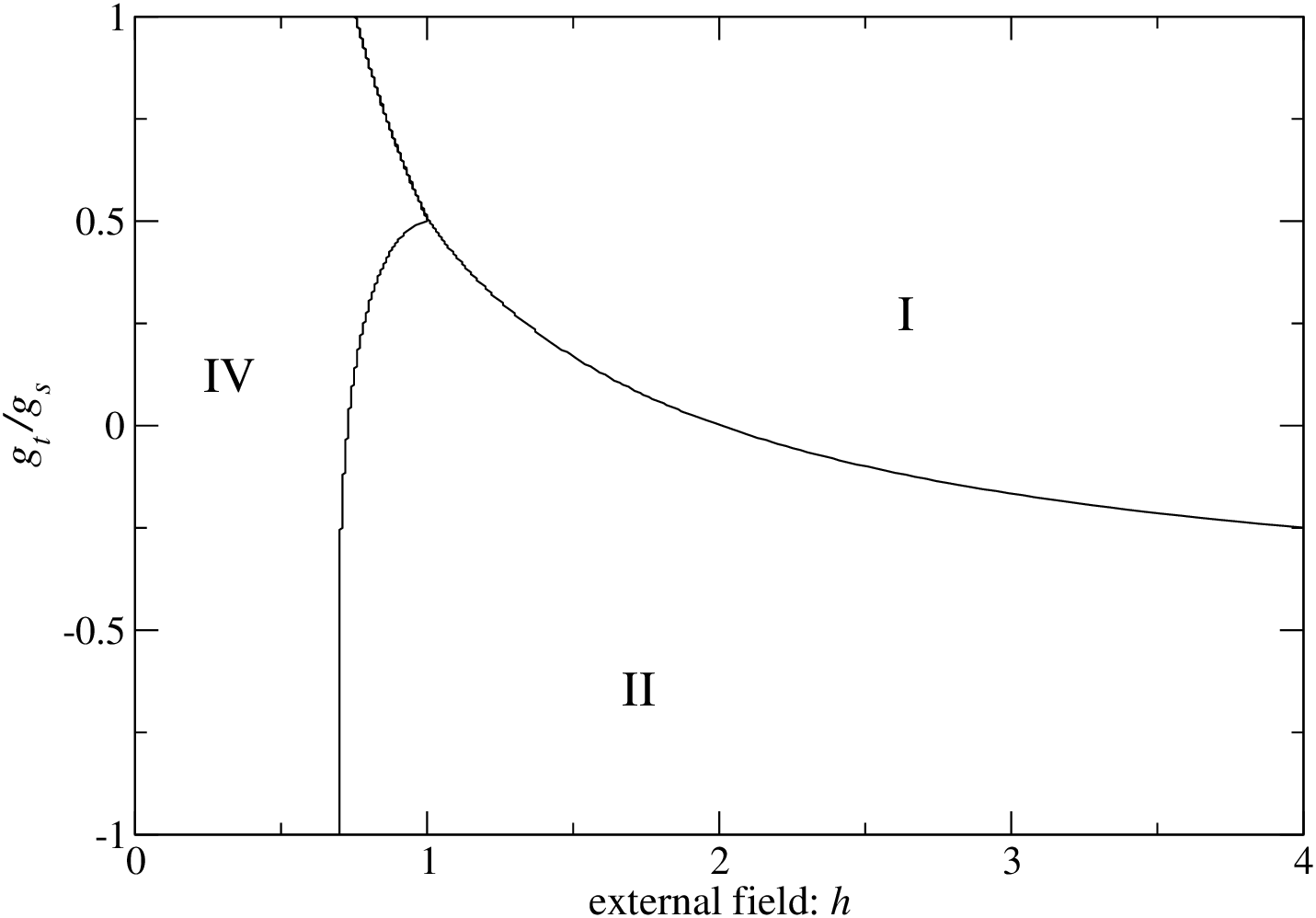}
\caption{\label{suheis_FIGURE_phase3} The phase diagram of
$g_t/g_s$versus $h$ with $g_u=g_s+g_t$ and residual symmetry
$U(1)\times U(1)\times SU(2)$}
\end{figure}

The magnetization curves for  different $g_t$ is plotted in Fig.
\ref{suheis_FIGURE_magn3}, and the phase diagram in terms of
$g_t/g_s$ versus $h$ is given in Fig. \ref{suheis_FIGURE_phase3}.
On the one hand, because the eigenvalue of $T^z$ equals that of
$U^z$ for both states $|\underline \uparrow\rangle$ and
$|\overline\uparrow\rangle$, the flipping process occurred in
phase II contributes to both magnetization of  $T^z$ and $U^z$
equivalently. On the other hand, the flipping from
$|\underline\downarrow\rangle$ and $|\overline \downarrow\rangle$
occurs simultaneously  in the phase IV due to the SU(2) symmetry.
As a results, the magnetization of $T^z$ and $U^z$ are expected to
be the same in the whole process, which can be seen from our
numerical calculation in Fig. \ref{suheis_FIGURE_magn3}.

For $g_t/g_s<0.5$, there exists three distinct phases, denoted by
IV, II and I respectively according to the number of the rows of
Young tableau. The SU(2) symmetry makes the states
$|\underline\downarrow\rangle$ and $|\overline\downarrow\rangle$
flip simultaneously when the external field increases. This makes
the four-row Young tableau turn to two-row Young tableau directly,
hence the phase III labelled by three-row Young tableau will not
take place. The boundary between phase IV and phase II is
determined from $\zeta(0)=0$ and $\xi(0)=0$ together,
\begin{eqnarray}
g_t/g_s=\frac{1}{2}+\frac{1}{2h}K_{1/2}(0)*\varepsilon(0),
\label{eq:phase3_line1}
\end{eqnarray}
where $\varepsilon$ can be computed from eqs. (\ref{eq:thermodeq})
numerically. For sufficient large external field, all $S$, $T$ and
$U$  are frozen to the $z$-direction, which brings about the
occurrence of phase I. The boundary between phase I and phase II
is determined by $\varepsilon(0)=0$, i.e.,
\begin{eqnarray}
g_t/g_s=\frac{1}{h}-\frac{1}{2}. \label{eq:phase3_line2}
\end{eqnarray}
This can also be derived from the competition between the states
related to the Young tableau $[N-1,1]$ and [$N$].

The asymptotic behavior at large $h$ is $g_t/g_s=-1/2$, which
implies that the phase I will never occur as long as
$g_t/g_s<-1/2$. The magnetization process in the region
$-1/2<g_t/g_s<1/2$  can be illustrated by the following evolution
of Young tableau,
\begin{center}
\setlength{\unitlength}{3mm}
\begin{picture}(18,5)(3,0)\linethickness{0.6pt}
\put(-1, 3.35){\vector(0,1){0.7}} \put(-1.2,3.35){\line(1,0){0.4}}
\put(-1, 2.15){\vector(0,1){0.7}} \put(-1.2,2.85){\line(1,0){0.4}}
\put(-1, 1.85){\vector(0,-1){0.7}}\put(-1.2,1.15){\line(1,0){0.4}}
\put(-1, 0.65){\vector(0,-1){0.7}}
\put(-1.2,0.65){\line(1,0){0.4}}

\put(0,1){\line(1,0){1}}
 \put(0,2){\line(1,0){1}}
 \put(0,3){\line(1,0){1}}
 \put(0,0){\line(1,0){1}}
  \put(0,4){\line(1,0){1}}
   \put(2.5,0){\line(1,0){1}}
  \put(2.5,4){\line(1,0){1}}
 \put(1.1,0){\line(1,0){0.1}} \put(1.4,0){\line(1,0){0.1}}  \put(1.7,0){\line(1,0){0.1}}
 \put(2.0,0){\line(1,0){0.1}} \put(2.3,0){\line(1,0){0.1}}  \put(2.6,0){\line(1,0){0.1}}
 \put(2.9,0){\line(1,0){0.1}}
 \put(1.1,1){\line(1,0){0.1}} \put(1.4,1){\line(1,0){0.1}}  \put(1.7,1){\line(1,0){0.1}}
 \put(2.0,1){\line(1,0){0.1}} \put(2.3,1){\line(1,0){0.1}}  \put(2.6,1){\line(1,0){0.1}}
 \put(2.9,1){\line(1,0){0.1}}
 \put(1.1,2){\line(1,0){0.1}} \put(1.4,2){\line(1,0){0.1}}  \put(1.7,2){\line(1,0){0.1}}
 \put(2.0,2){\line(1,0){0.1}} \put(2.3,2){\line(1,0){0.1}}  \put(2.6,2){\line(1,0){0.1}}
 \put(2.9,2){\line(1,0){0.1}}
 \put(1.1,3){\line(1,0){0.1}} \put(1.4,3){\line(1,0){0.1}}  \put(1.7,3){\line(1,0){0.1}}
 \put(2.0,3){\line(1,0){0.1}} \put(2.3,3){\line(1,0){0.1}}  \put(2.6,3){\line(1,0){0.1}}
 \put(2.9,3){\line(1,0){0.1}}
 \put(1.1,4){\line(1,0){0.1}} \put(1.4,4){\line(1,0){0.1}}  \put(1.7,4){\line(1,0){0.1}}
 \put(2.0,4){\line(1,0){0.1}} \put(2.3,4){\line(1,0){0.1}}  \put(2.6,4){\line(1,0){0.1}}
 \put(2.9,4){\line(1,0){0.1}}
  \put(2.5,1){\line(1,0){1}}
   \put(2.5,2){\line(1,0){1}}
    \put(2.5,3){\line(1,0){1}}
 \put(3.5,0){\line(0,1){4}}
  \put(2.5,0){\line(0,1){4}}
   \put(1,0){\line(0,1){4}}
   \put(0,0){\line(0,1){4}}

\put(4,3.5){\vector(1,0){1}}

\put(5.5,0){\line(0,1){4}} \put(6.5,0){\line(0,1){4}}
 \put(8,0){\line(0,1){4}} \put(9,2){\line(0,1){2}}
    \put(10.5,2){\line(0,1){2}} \put(11.5,3){\line(0,1){1}}
    \put(13,3){\line(0,1){1}}

\put(5.5,0){\line(1,0){1}} \put(8,2){\line(1,0){1}}
\put(10.5,3){\line(1,0){1}} \put(5.5,4){\line(1,0){1}}
\put(8,4){\line(1,0){1}} \put(10.5,4){\line(1,0){1}}

\put(5.5,1){\line(1,0){1}}
   \put(5.5,2){\line(1,0){1}}
    \put(5.5,3){\line(1,0){1}}
\put(8,3){\line(1,0){1}}

 \put(6.6,4){\line(1,0){0.1}} \put(6.9,4){\line(1,0){0.1}}  \put(7.2,4){\line(1,0){0.1}}
 \put(7.5,4){\line(1,0){0.1}} \put(7.8,4){\line(1,0){0.1}}
 \put(6.6,3){\line(1,0){0.1}} \put(6.9,3){\line(1,0){0.1}}  \put(7.2,3){\line(1,0){0.1}}
 \put(7.5,3){\line(1,0){0.1}} \put(7.8,3){\line(1,0){0.1}}
 \put(6.6,2){\line(1,0){0.1}} \put(6.9,2){\line(1,0){0.1}}  \put(7.2,2){\line(1,0){0.1}}
 \put(7.5,2){\line(1,0){0.1}} \put(7.8,2){\line(1,0){0.1}}
 \put(6.6,1){\line(1,0){0.1}} \put(6.9,1){\line(1,0){0.1}}  \put(7.2,1){\line(1,0){0.1}}
 \put(7.5,1){\line(1,0){0.1}} \put(7.8,1){\line(1,0){0.1}}
 \put(6.6,0){\line(1,0){0.1}} \put(6.9,0){\line(1,0){0.1}}  \put(7.2,0){\line(1,0){0.1}}
 \put(7.5,0){\line(1,0){0.1}} \put(7.8,0){\line(1,0){0.1}}

 \put(9.1,4){\line(1,0){0.1}} \put(9.4,4){\line(1,0){0.1}}  \put(9.7,4){\line(1,0){0.1}}
 \put(10,4){\line(1,0){0.1}} \put(10.3,4){\line(1,0){0.1}}
 \put(9.1,3){\line(1,0){0.1}} \put(9.4,3){\line(1,0){0.1}}  \put(9.7,3){\line(1,0){0.1}}
 \put(10,3){\line(1,0){0.1}} \put(10.3,3){\line(1,0){0.1}}
 \put(9.1,2){\line(1,0){0.1}} \put(9.4,2){\line(1,0){0.1}}  \put(9.7,2){\line(1,0){0.1}}
 \put(10,2){\line(1,0){0.1}} \put(10.3,2){\line(1,0){0.1}}

 \put(11.6,3){\line(1,0){0.1}} \put(11.9,3){\line(1,0){0.1}} \put(12.2,3){\line(1,0){0.1}}
 \put(12.5,3){\line(1,0){0.1}} \put(12.8,3){\line(1,0){0.1}}
 \put(11.6,4){\line(1,0){0.1}} \put(11.9,4){\line(1,0){0.1}} \put(12.2,4){\line(1,0){0.1}}
 \put(12.5,4){\line(1,0){0.1}} \put(12.8,4){\line(1,0){0.1}}

\put(13.5,3.5){\vector(1,0){1}}

\put(15,2){\line(0,1){2}} \put(16,2){\line(0,1){2}}
\put(17.5,2){\line(0,1){2}} \put(18.5,3){\line(0,1){1}}
\put(20,3){\line(0,1){1}}

\put(15,2){\line(1,0){1}} \put(17.5,3){\line(1,0){1}}
\put(15,4){\line(1,0){1}} \put(17.5,4){\line(1,0){1}}
\put(15,3){\line(1,0){1}}

\put(16.1,4){\line(1,0){0.1}} \put(16.4,4){\line(1,0){0.1}}
\put(16.7,4){\line(1,0){0.1}} \put(17,4){\line(1,0){0.1}}
\put(17.3,4){\line(1,0){0.1}} \put(16.1,3){\line(1,0){0.1}}
\put(16.4,3){\line(1,0){0.1}} \put(16.7,3){\line(1,0){0.1}}
\put(17,3){\line(1,0){0.1}} \put(17.3,3){\line(1,0){0.1}}

\put(16.1,2){\line(1,0){0.1}} \put(16.4,2){\line(1,0){0.1}}
\put(16.7,2){\line(1,0){0.1}} \put(17,2){\line(1,0){0.1}}
\put(17.3,2){\line(1,0){0.1}}

\put(18.6,3){\line(1,0){0.1}} \put(18.9,3){\line(1,0){0.1}}
\put(19.2,3){\line(1,0){0.1}} \put(19.5,3){\line(1,0){0.1}}
\put(19.8,3){\line(1,0){0.1}}

\put(18.6,4){\line(1,0){0.1}} \put(18.9,4){\line(1,0){0.1}}
\put(19.2,4){\line(1,0){0.1}} \put(19.5,4){\line(1,0){0.1}}
\put(19.8,4){\line(1,0){0.1}}

\put(20.5,3.5){\vector(1,0){1}}

\put(22,3){\line(0,1){1}} \put(23,3){\line(0,1){1}}
\put(24.5,3){\line(0,1){1}} \put(25.5,3){\line(0,1){1}}
\put(22,3){\line(1,0){1}} \put(22,4){\line(1,0){1}}
\put(24.5,3){\line(1,0){1}} \put(24.5,4){\line(1,0){1}}
\put(23.1,3){\line(1,0){0.1}} \put(23.4,3){\line(1,0){0.1}}
\put(23.7,3){\line(1,0){0.1}} \put(24.0,3){\line(1,0){0.1}}
\put(24.3,3){\line(1,0){0.1}} \put(23.1,4){\line(1,0){0.1}}
\put(23.4,4){\line(1,0){0.1}} \put(23.7,4){\line(1,0){0.1}}
\put(24.0,4){\line(1,0){0.1}} \put(24.3,4){\line(1,0){0.1}}
\end{picture}
\end{center}
The boundary between  phase IV and phase I is determined by
$\varepsilon(0)=0$, $\zeta(0)=0$ and $\xi(0)$ together,
\begin{eqnarray}
g_t/g_s=\frac{3}{2h}-1.
\label{eq:phase3 line3}
\end{eqnarray}
The common solution of
Eqs. (\ref{eq:phase3_line1}-\ref{eq:phase3
line3}) gives $h=1$ and $g_t/g_s =1/2$, which is a
three-phase-coexist point.

\subsection{Residual U(1)$\times$ SU(2) $\times$ U(1) symmetry}

If the external field along the direction of the second simple
root is quenched but those along the other directions are kept, we
will have symmetry breaking down to U(1)$\times$ SU(2) $\times$
U(1). Such kind of symmetry breaking is caused by Zeeman term of
the following magnetization
\begin{eqnarray}
{\cal M}^z =(g_t+g_u)O_1+(g_t-g_u)O_3.
\end{eqnarray}

\begin{figure*}
\includegraphics[width=15cm]{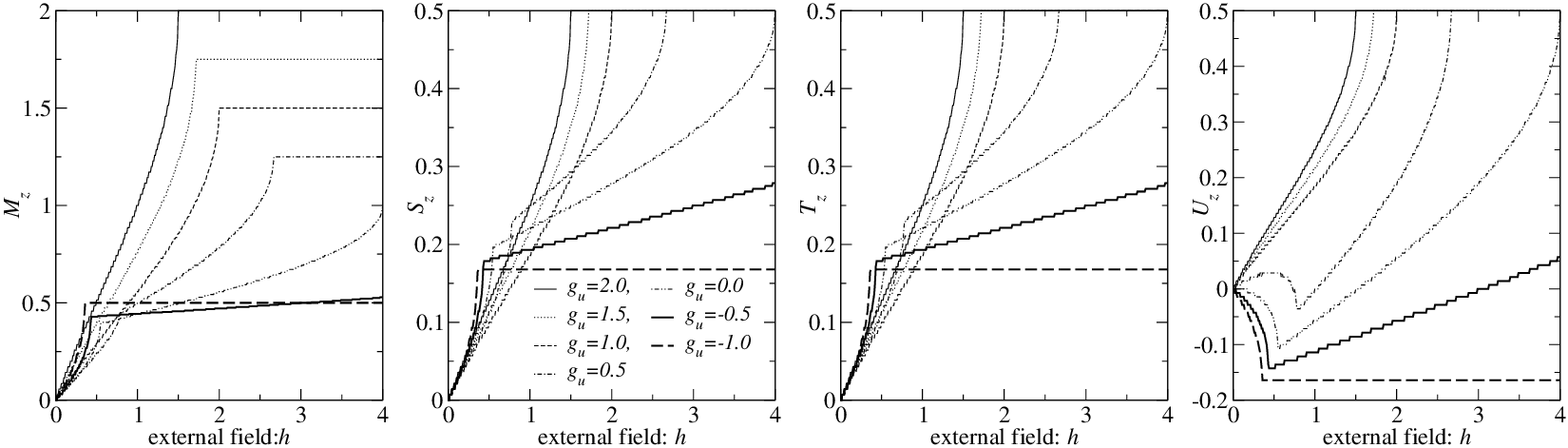}
\caption{\label{suheis_FIGURE_magn4} The magnetization ${\cal M}^z
, S_z, T_z, U_z$ of the system with $g_s=0, g_t=2$, and $g_u=2.0,
1.5, 1.0, 0.5, 0.0, -0.5, -1.0$.\\\\}
\end{figure*}

\begin{figure}
\includegraphics[width=7cm]{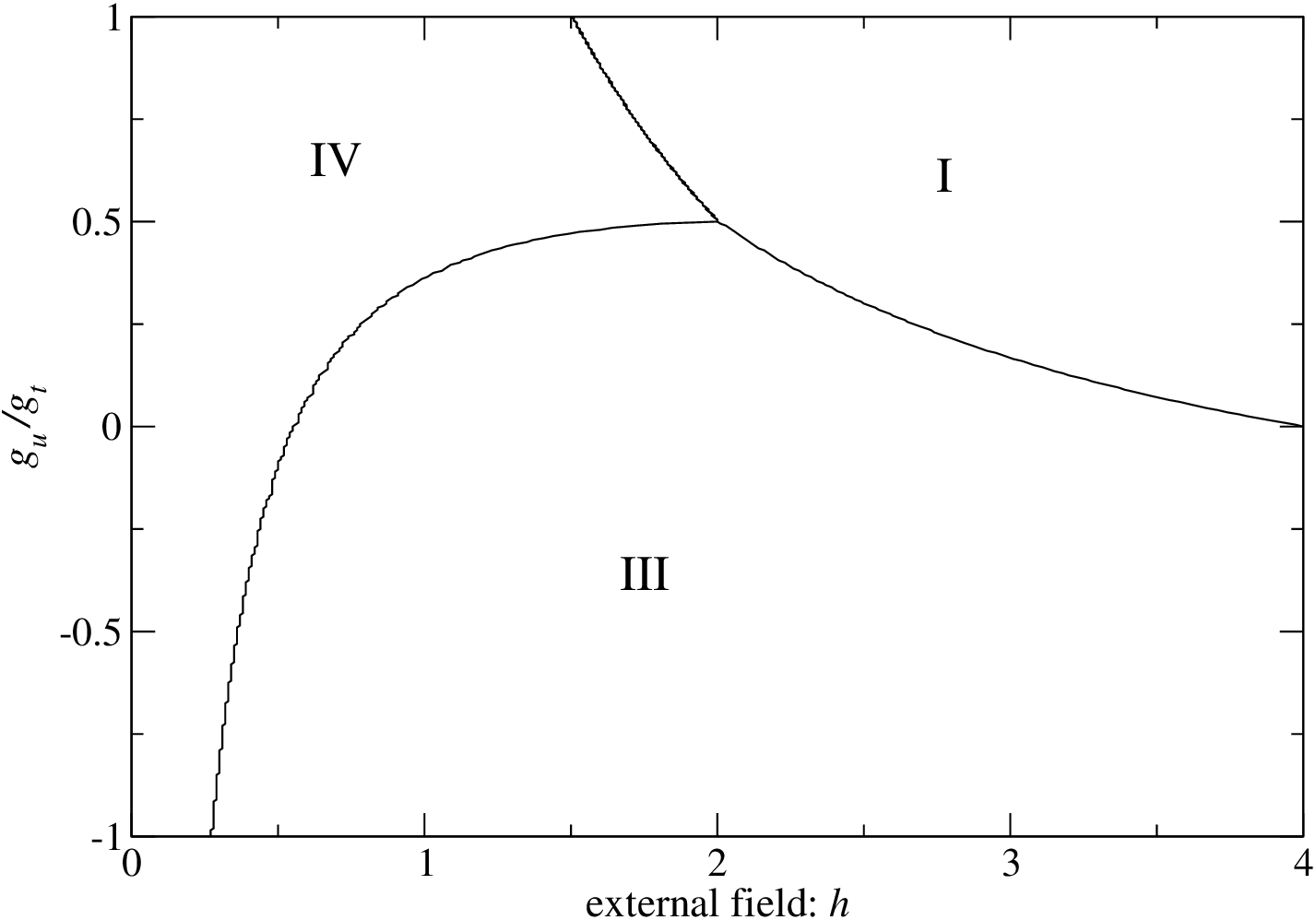}
\caption{\label{suheis_FIGURE_phase4} The phase diagram of
$g_u/g_t$ versus $h$ with $g_s=0$ and residual symmetry
$U(1)\times SU(2) \times U(1)$}
\end{figure}

The magnetization curves with different $g_u$ is plotted in Fig.
\ref{suheis_FIGURE_magn4}, and the phase diagram in terms of
$g_u/g_t$ versus $h$ is given in Fig. \ref{suheis_FIGURE_phase4}.
For $g_u/g_t<1/2$, there exists three phases denoted by IV, III
and I respectively.

The boundary between phases IV and III is determined from $\xi=0$,
which can be solved numerically. For sufficient large external
field, the magnetization is saturated reaching phase I. This phase
transition occurs at
\begin{equation}
g_u/g_t=\frac{2}{h}-\frac{1}{2}
\end{equation}
Thus if $g_u/g_t<-1/2$, the phase I will never occur regardless of
the magnitude of external field. So in the region
$-1/2<g_u/g_t<1/2$, the magnetization process can be illustrated
by Young tableau
\begin{center}
\setlength{\unitlength}{3mm}
\begin{picture}(18,5)(3,0)\linethickness{0.6pt}
\put(-1, 3.35){\vector(0,1){0.7}} \put(-1.2,3.35){\line(1,0){0.4}}
\put(-1, 2.15){\vector(0,1){0.7}} \put(-1.2,2.85){\line(1,0){0.4}}
\put(-1, 1.85){\vector(0,-1){0.7}}\put(-1.2,1.15){\line(1,0){0.4}}
\put(-1, 0.65){\vector(0,-1){0.7}}
\put(-1.2,0.65){\line(1,0){0.4}}

\put(0,1){\line(1,0){1}}
 \put(0,2){\line(1,0){1}}
 \put(0,3){\line(1,0){1}}
 \put(0,0){\line(1,0){1}}
  \put(0,4){\line(1,0){1}}
   \put(2.5,0){\line(1,0){1}}
  \put(2.5,4){\line(1,0){1}}
 \put(1.1,0){\line(1,0){0.1}} \put(1.4,0){\line(1,0){0.1}}  \put(1.7,0){\line(1,0){0.1}}
 \put(2.0,0){\line(1,0){0.1}} \put(2.3,0){\line(1,0){0.1}}  \put(2.6,0){\line(1,0){0.1}}
 \put(2.9,0){\line(1,0){0.1}}
 \put(1.1,1){\line(1,0){0.1}} \put(1.4,1){\line(1,0){0.1}}  \put(1.7,1){\line(1,0){0.1}}
 \put(2.0,1){\line(1,0){0.1}} \put(2.3,1){\line(1,0){0.1}}  \put(2.6,1){\line(1,0){0.1}}
 \put(2.9,1){\line(1,0){0.1}}
 \put(1.1,2){\line(1,0){0.1}} \put(1.4,2){\line(1,0){0.1}}  \put(1.7,2){\line(1,0){0.1}}
 \put(2.0,2){\line(1,0){0.1}} \put(2.3,2){\line(1,0){0.1}}  \put(2.6,2){\line(1,0){0.1}}
 \put(2.9,2){\line(1,0){0.1}}
 \put(1.1,3){\line(1,0){0.1}} \put(1.4,3){\line(1,0){0.1}}  \put(1.7,3){\line(1,0){0.1}}
 \put(2.0,3){\line(1,0){0.1}} \put(2.3,3){\line(1,0){0.1}}  \put(2.6,3){\line(1,0){0.1}}
 \put(2.9,3){\line(1,0){0.1}}
 \put(1.1,4){\line(1,0){0.1}} \put(1.4,4){\line(1,0){0.1}}  \put(1.7,4){\line(1,0){0.1}}
 \put(2.0,4){\line(1,0){0.1}} \put(2.3,4){\line(1,0){0.1}}  \put(2.6,4){\line(1,0){0.1}}
 \put(2.9,4){\line(1,0){0.1}}
  \put(2.5,1){\line(1,0){1}}
   \put(2.5,2){\line(1,0){1}}
    \put(2.5,3){\line(1,0){1}}
 \put(3.5,0){\line(0,1){4}}
  \put(2.5,0){\line(0,1){4}}
   \put(1,0){\line(0,1){4}}
   \put(0,0){\line(0,1){4}}

\put(4,3.5){\vector(1,0){1}}

\put(5.5,0){\line(0,1){4}} \put(6.5,0){\line(0,1){4}}
 \put(8,0){\line(0,1){4}} \put(9,1){\line(0,1){3}}
    \put(10.5,1){\line(0,1){3}} \put(11.5,3){\line(0,1){1}}
    \put(13,3){\line(0,1){1}}

\put(5.5,0){\line(1,0){1}} \put(8,2){\line(1,0){1}}
\put(10.5,3){\line(1,0){1}} \put(5.5,4){\line(1,0){1}}
\put(8,4){\line(1,0){1}} \put(10.5,4){\line(1,0){1}}

\put(5.5,1){\line(1,0){1}}
   \put(5.5,2){\line(1,0){1}}
    \put(5.5,3){\line(1,0){1}}
\put(8,3){\line(1,0){1}}

 \put(6.6,4){\line(1,0){0.1}} \put(6.9,4){\line(1,0){0.1}}  \put(7.2,4){\line(1,0){0.1}}
 \put(7.5,4){\line(1,0){0.1}} \put(7.8,4){\line(1,0){0.1}}
 \put(6.6,3){\line(1,0){0.1}} \put(6.9,3){\line(1,0){0.1}}  \put(7.2,3){\line(1,0){0.1}}
 \put(7.5,3){\line(1,0){0.1}} \put(7.8,3){\line(1,0){0.1}}
 \put(6.6,2){\line(1,0){0.1}} \put(6.9,2){\line(1,0){0.1}}  \put(7.2,2){\line(1,0){0.1}}
 \put(7.5,2){\line(1,0){0.1}} \put(7.8,2){\line(1,0){0.1}}
 \put(6.6,1){\line(1,0){0.1}} \put(6.9,1){\line(1,0){0.1}}  \put(7.2,1){\line(1,0){0.1}}
 \put(7.5,1){\line(1,0){0.1}} \put(7.8,1){\line(1,0){0.1}}
 \put(6.6,0){\line(1,0){0.1}} \put(6.9,0){\line(1,0){0.1}}  \put(7.2,0){\line(1,0){0.1}}
 \put(7.5,0){\line(1,0){0.1}} \put(7.8,0){\line(1,0){0.1}}

 \put(9.1,4){\line(1,0){0.1}} \put(9.4,4){\line(1,0){0.1}}  \put(9.7,4){\line(1,0){0.1}}
 \put(10,4){\line(1,0){0.1}} \put(10.3,4){\line(1,0){0.1}}
 \put(9.1,3){\line(1,0){0.1}} \put(9.4,3){\line(1,0){0.1}}  \put(9.7,3){\line(1,0){0.1}}
 \put(10,3){\line(1,0){0.1}} \put(10.3,3){\line(1,0){0.1}}
 \put(9.1,2){\line(1,0){0.1}} \put(9.4,2){\line(1,0){0.1}}  \put(9.7,2){\line(1,0){0.1}}
 \put(10,2){\line(1,0){0.1}} \put(10.3,2){\line(1,0){0.1}}
 \put(9.1,1){\line(1,0){0.1}} \put(9.4,1){\line(1,0){0.1}}  \put(9.7,1){\line(1,0){0.1}}
 \put(10,1){\line(1,0){0.1}} \put(10.3,1){\line(1,0){0.1}}
 \put(8,1){\line(1,0){1}}
 \put(11.6,3){\line(1,0){0.1}} \put(11.9,3){\line(1,0){0.1}} \put(12.2,3){\line(1,0){0.1}}
 \put(12.5,3){\line(1,0){0.1}} \put(12.8,3){\line(1,0){0.1}}
 \put(11.6,4){\line(1,0){0.1}} \put(11.9,4){\line(1,0){0.1}} \put(12.2,4){\line(1,0){0.1}}
 \put(12.5,4){\line(1,0){0.1}} \put(12.8,4){\line(1,0){0.1}}

\put(13.5,3.5){\vector(1,0){1}}

\put(15,1){\line(0,1){3}} \put(16,1){\line(0,1){3}}
\put(17.5,1){\line(0,1){3}} \put(18.5,3){\line(0,1){1}}
\put(20,3){\line(0,1){1}}

\put(15,1){\line(1,0){1}} \put(17.5,3){\line(1,0){1}}
\put(15,4){\line(1,0){1}} \put(17.5,4){\line(1,0){1}}
\put(15,3){\line(1,0){1}} \put(15,2){\line(1,0){1}}

\put(16.1,3){\line(1,0){0.1}} \put(16.4,3){\line(1,0){0.1}}
\put(16.7,3){\line(1,0){0.1}} \put(17,3){\line(1,0){0.1}}
\put(17.3,3){\line(1,0){0.1}} \put(16.1,2){\line(1,0){0.1}}
\put(16.4,2){\line(1,0){0.1}} \put(16.7,2){\line(1,0){0.1}}
\put(17,2){\line(1,0){0.1}} \put(17.3,2){\line(1,0){0.1}}
\put(16.1,4){\line(1,0){0.1}} \put(16.4,4){\line(1,0){0.1}}
\put(16.7,4){\line(1,0){0.1}} \put(17,4){\line(1,0){0.1}}
\put(17.3,4){\line(1,0){0.1}} \put(16.1,1){\line(1,0){0.1}}
\put(16.4,1){\line(1,0){0.1}} \put(16.7,1){\line(1,0){0.1}}
\put(17,1){\line(1,0){0.1}} \put(17.3,1){\line(1,0){0.1}}

\put(18.6,4){\line(1,0){0.1}} \put(18.9,4){\line(1,0){0.1}}
\put(19.2,4){\line(1,0){0.1}} \put(19.5,4){\line(1,0){0.1}}
\put(19.8,4){\line(1,0){0.1}} \put(18.6,3){\line(1,0){0.1}}
\put(18.9,3){\line(1,0){0.1}} \put(19.2,3){\line(1,0){0.1}}
\put(19.5,3){\line(1,0){0.1}} \put(19.8,3){\line(1,0){0.1}}

 \put(20.5,3.5){\vector(1,0){1}}

\put(22,3){\line(0,1){1}} \put(23,3){\line(0,1){1}}
\put(24.5,3){\line(0,1){1}} \put(25.5,3){\line(0,1){1}}
\put(22,3){\line(1,0){1}} \put(22,4){\line(1,0){1}}
\put(24.5,3){\line(1,0){1}} \put(24.5,4){\line(1,0){1}}

\put(23.1,3){\line(1,0){0.1}} \put(23.4,3){\line(1,0){0.1}}
\put(23.7,3){\line(1,0){0.1}} \put(24.0,3){\line(1,0){0.1}}
\put(24.3,3){\line(1,0){0.1}} \put(23.1,4){\line(1,0){0.1}}
\put(23.4,4){\line(1,0){0.1}} \put(23.7,4){\line(1,0){0.1}}
\put(24.0,4){\line(1,0){0.1}} \put(24.3,4){\line(1,0){0.1}}
\end{picture}
\end{center}

In phase III, the length of the second row and that of the third
row in their corresponding Young tableau is always equal due to
the SU(2) symmetry. Thus the probability of pure spin-flipping
$|\overline\downarrow\rangle \rightarrow
|\overline\uparrow\rangle$ and pure orbital-flipping
$|\overline\downarrow\rangle \rightarrow
|\underline\downarrow\rangle$ is the same. Additionally, the
process $|\overline\downarrow\rangle \rightarrow
|\underline\uparrow\rangle$ contributes the same for the
magnetization of $S^z$ and $T^z$. These properties result in the
same magnetization curves of $S^z$ and $T^z$ shown in Fig.
\ref{suheis_FIGURE_magn4}. Flipping over the state $|\overline
\downarrow\rangle$ which has positive eigenvalue of $U^z$ brings
about a negative magnetization of $U^z$ in phase IV.

The phase I will never occur for $g_u/g_t<-1/2$, as is similar to
the case of SU(3)$\times$U(1). Actually, it recovers the case of
residual SU(3)$\times$U(1) symmetry at $g_u/g_t=-1$,

If $g_u/g_t>1/2$,  we can see from Fig. \ref{suheis_FIGURE_phase4}
that the phase IV transits into phase I directly along with the
increase of external field. The boundary  which separates these
two phase is determined by
\begin{eqnarray}
g_u/g_t=\frac{3}{h}-1.
\end{eqnarray}
Obviously, the point at $g_u/g_t=0.5$ and $h=2$ is a
three-phase-coexist point. The magnetization properties in the
region of $g_u/g_t>1/2$ is similar to the case of
U(1)$\times$SU(3), particulary, a larger U(1)$\times$SU(3)
symmetry remains for $g_u/g_t=1$.

\subsection{Residual SU(2)$\times$ U(1) $\times$ U(1) symmetry }

The third case of two-parameter symmetry breaking is produced by
the Zeeman term with the restriction to the the Land\'e g factor
$g_u=-g_s-g_t$. The magnetization now reads
\begin{eqnarray}
{\cal M}^z =2 g_s O^z_2 +2(g_s+g_t)O^z_3.
\end{eqnarray}
which breaks the SU(4) symmetry down to SU(2)$\times$ U(1)
$\times$ U(1).

\begin{figure*}
\includegraphics[width=15cm]{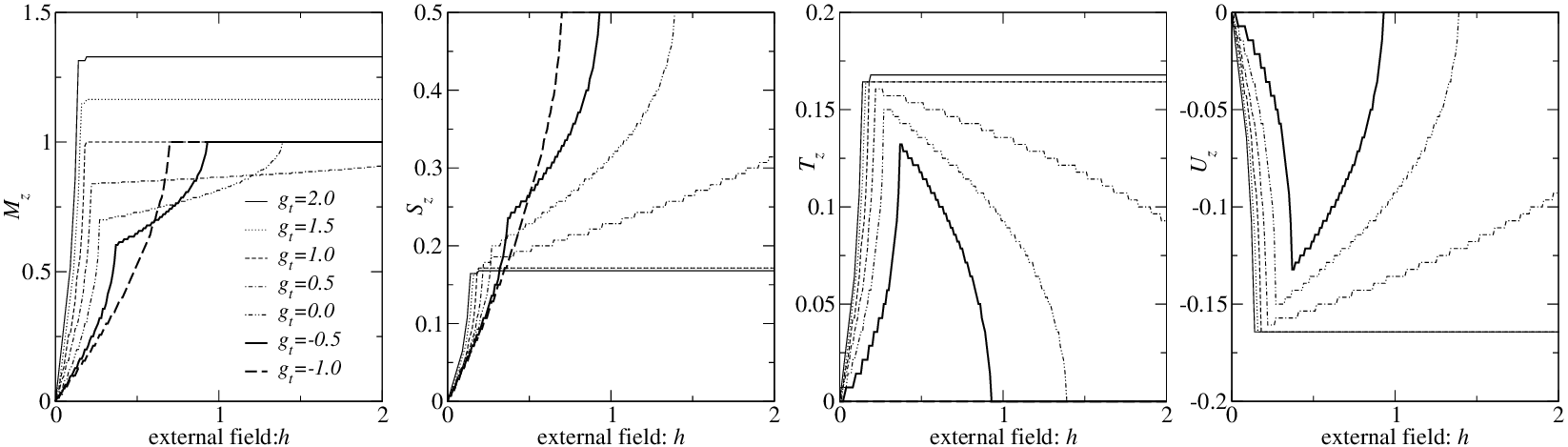}
\caption{\label{suheis_FIGURE_magn5} The magnetization ${\cal M}^z
, S_z, T_z, U_z$ of the system with $g_s=2, g_u=-g_s-g_t$, and
$g_t=2.0, 1.5, 1.0, 0.5, 0.0, -0.5, -1.0$.\\\\}
\end{figure*}

\begin{figure}
\includegraphics[width=7cm]{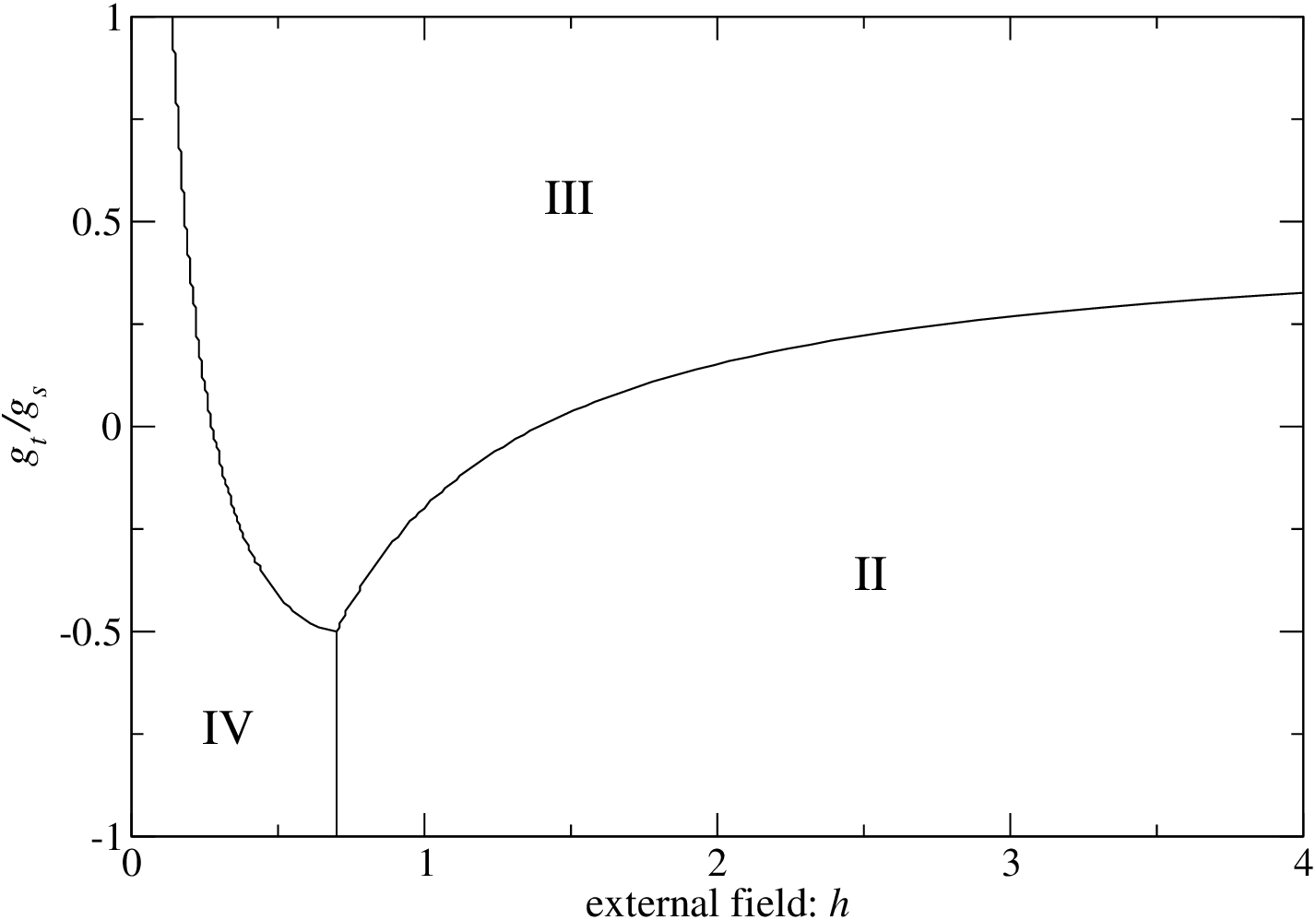}
\caption{\label{suheis_FIGURE_phase5} The phase diagram of
$g_t/g_s$ versus $h$ with $g_u=-g_s-g_t$. and residual symmetry
$SU(2)\times U(1) \times U(1)$}
\end{figure}

The magnetization curves for different $g_t/g_s$ is plotted in
Fig. \ref{suheis_FIGURE_magn5}, and phase diagram in terms of
$g_t/g_s$ versus $h$ is given in Fig. \ref{suheis_FIGURE_phase5}.
In the region of $g_t/g_s>-1/2$, there exist three phases denoted
by IV, III, and II respectively. The final state is characterized
by two rows Young tableau at sufficient large external field due
to the SU(2) symmetry. The boundary separates  phase III and II is
determined by $\zeta(0)=0$, which reads
\begin{eqnarray}
g_t/g_s=\frac{1}{2}- \frac{\ln 2}{h}.
\end{eqnarray}
Obviously if $g_t/g_s >1/2$, the phase II will never occur
regardless the magnitude of external field, and the magnetization
process is similar to the case of SU(3)$\times$U(1).

In the region $-1/2<g_t/g_s<1/2$, the magnetization process can be
illustrated by the following Young tableau,
\begin{center}
\setlength{\unitlength}{3mm}
\begin{picture}(18,5)(3,0)\linethickness{0.6pt}
\put(-1, 3.35){\vector(0,1){0.7}} \put(-1.2,3.35){\line(1,0){0.4}}
\put(-1, 2.15){\vector(0,1){0.7}} \put(-1.2,2.85){\line(1,0){0.4}}
\put(-1, 1.85){\vector(0,-1){0.7}}\put(-1.2,1.15){\line(1,0){0.4}}
\put(-1, 0.65){\vector(0,-1){0.7}}
\put(-1.2,0.65){\line(1,0){0.4}}

\put(0,1){\line(1,0){1}}
 \put(0,2){\line(1,0){1}}
 \put(0,3){\line(1,0){1}}
 \put(0,0){\line(1,0){1}}
  \put(0,4){\line(1,0){1}}
   \put(2.5,0){\line(1,0){1}}
  \put(2.5,4){\line(1,0){1}}
 \put(1.1,0){\line(1,0){0.1}} \put(1.4,0){\line(1,0){0.1}}  \put(1.7,0){\line(1,0){0.1}}
 \put(2.0,0){\line(1,0){0.1}} \put(2.3,0){\line(1,0){0.1}}  \put(2.6,0){\line(1,0){0.1}}
 \put(2.9,0){\line(1,0){0.1}}
 \put(1.1,1){\line(1,0){0.1}} \put(1.4,1){\line(1,0){0.1}}  \put(1.7,1){\line(1,0){0.1}}
 \put(2.0,1){\line(1,0){0.1}} \put(2.3,1){\line(1,0){0.1}}  \put(2.6,1){\line(1,0){0.1}}
 \put(2.9,1){\line(1,0){0.1}}
 \put(1.1,2){\line(1,0){0.1}} \put(1.4,2){\line(1,0){0.1}}  \put(1.7,2){\line(1,0){0.1}}
 \put(2.0,2){\line(1,0){0.1}} \put(2.3,2){\line(1,0){0.1}}  \put(2.6,2){\line(1,0){0.1}}
 \put(2.9,2){\line(1,0){0.1}}
 \put(1.1,3){\line(1,0){0.1}} \put(1.4,3){\line(1,0){0.1}}  \put(1.7,3){\line(1,0){0.1}}
 \put(2.0,3){\line(1,0){0.1}} \put(2.3,3){\line(1,0){0.1}}  \put(2.6,3){\line(1,0){0.1}}
 \put(2.9,3){\line(1,0){0.1}}
 \put(1.1,4){\line(1,0){0.1}} \put(1.4,4){\line(1,0){0.1}}  \put(1.7,4){\line(1,0){0.1}}
 \put(2.0,4){\line(1,0){0.1}} \put(2.3,4){\line(1,0){0.1}}  \put(2.6,4){\line(1,0){0.1}}
 \put(2.9,4){\line(1,0){0.1}}
  \put(2.5,1){\line(1,0){1}}
   \put(2.5,2){\line(1,0){1}}
    \put(2.5,3){\line(1,0){1}}
 \put(3.5,0){\line(0,1){4}}
  \put(2.5,0){\line(0,1){4}}
   \put(1,0){\line(0,1){4}}
   \put(0,0){\line(0,1){4}}

\put(4,3.5){\vector(1,0){1}}

\put(5.5,0){\line(0,1){4}} \put(6.5,0){\line(0,1){4}}
 \put(8,0){\line(0,1){4}} \put(9,1){\line(0,1){3}}
    \put(10.5,1){\line(0,1){3}} \put(11.5,2){\line(0,1){2}}
    \put(13,2){\line(0,1){2}} \put(10.5,2){\line(1,0){1}}

\put(5.5,0){\line(1,0){1}} \put(8,2){\line(1,0){1}}
\put(10.5,3){\line(1,0){1}} \put(5.5,4){\line(1,0){1}}
\put(8,4){\line(1,0){1}} \put(10.5,4){\line(1,0){1}}

\put(5.5,1){\line(1,0){1}}
   \put(5.5,2){\line(1,0){1}}
    \put(5.5,3){\line(1,0){1}}
\put(8,3){\line(1,0){1}}

 \put(6.6,4){\line(1,0){0.1}} \put(6.9,4){\line(1,0){0.1}}  \put(7.2,4){\line(1,0){0.1}}
 \put(7.5,4){\line(1,0){0.1}} \put(7.8,4){\line(1,0){0.1}}
 \put(6.6,3){\line(1,0){0.1}} \put(6.9,3){\line(1,0){0.1}}  \put(7.2,3){\line(1,0){0.1}}
 \put(7.5,3){\line(1,0){0.1}} \put(7.8,3){\line(1,0){0.1}}
 \put(6.6,2){\line(1,0){0.1}} \put(6.9,2){\line(1,0){0.1}}  \put(7.2,2){\line(1,0){0.1}}
 \put(7.5,2){\line(1,0){0.1}} \put(7.8,2){\line(1,0){0.1}}
 \put(6.6,1){\line(1,0){0.1}} \put(6.9,1){\line(1,0){0.1}}  \put(7.2,1){\line(1,0){0.1}}
 \put(7.5,1){\line(1,0){0.1}} \put(7.8,1){\line(1,0){0.1}}
 \put(6.6,0){\line(1,0){0.1}} \put(6.9,0){\line(1,0){0.1}}  \put(7.2,0){\line(1,0){0.1}}
 \put(7.5,0){\line(1,0){0.1}} \put(7.8,0){\line(1,0){0.1}}

 \put(9.1,4){\line(1,0){0.1}} \put(9.4,4){\line(1,0){0.1}}  \put(9.7,4){\line(1,0){0.1}}
 \put(10,4){\line(1,0){0.1}} \put(10.3,4){\line(1,0){0.1}}
 \put(9.1,3){\line(1,0){0.1}} \put(9.4,3){\line(1,0){0.1}}  \put(9.7,3){\line(1,0){0.1}}
 \put(10,3){\line(1,0){0.1}} \put(10.3,3){\line(1,0){0.1}}
 \put(9.1,2){\line(1,0){0.1}} \put(9.4,2){\line(1,0){0.1}}  \put(9.7,2){\line(1,0){0.1}}
 \put(10,2){\line(1,0){0.1}} \put(10.3,2){\line(1,0){0.1}}
 \put(9.1,1){\line(1,0){0.1}} \put(9.4,1){\line(1,0){0.1}}  \put(9.7,1){\line(1,0){0.1}}
 \put(10,1){\line(1,0){0.1}} \put(10.3,1){\line(1,0){0.1}}
 \put(8,1){\line(1,0){1}}
 \put(11.6,3){\line(1,0){0.1}} \put(11.9,3){\line(1,0){0.1}} \put(12.2,3){\line(1,0){0.1}}
 \put(12.5,3){\line(1,0){0.1}} \put(12.8,3){\line(1,0){0.1}}
 \put(11.6,4){\line(1,0){0.1}} \put(11.9,4){\line(1,0){0.1}} \put(12.2,4){\line(1,0){0.1}}
 \put(12.5,4){\line(1,0){0.1}} \put(12.8,4){\line(1,0){0.1}}
 \put(11.6,2){\line(1,0){0.1}} \put(11.9,2){\line(1,0){0.1}} \put(12.2,2){\line(1,0){0.1}}
 \put(12.5,2){\line(1,0){0.1}} \put(12.8,2){\line(1,0){0.1}}
\put(13.5,3.5){\vector(1,0){1}}

\put(15,1){\line(0,1){3}} \put(16,1){\line(0,1){3}}
\put(17.5,1){\line(0,1){3}} \put(18.5,2){\line(0,1){2}}
\put(20,2){\line(0,1){2}} \put(17.5,2){\line(1,0){1}}

\put(15,1){\line(1,0){1}} \put(17.5,3){\line(1,0){1}}
\put(15,4){\line(1,0){1}} \put(17.5,4){\line(1,0){1}}
\put(15,3){\line(1,0){1}} \put(15,2){\line(1,0){1}}

\put(16.1,3){\line(1,0){0.1}} \put(16.4,3){\line(1,0){0.1}}
\put(16.7,3){\line(1,0){0.1}} \put(17,3){\line(1,0){0.1}}
\put(17.3,3){\line(1,0){0.1}} \put(16.1,2){\line(1,0){0.1}}
\put(16.4,2){\line(1,0){0.1}} \put(16.7,2){\line(1,0){0.1}}
\put(17,2){\line(1,0){0.1}} \put(17.3,2){\line(1,0){0.1}}
\put(16.1,4){\line(1,0){0.1}} \put(16.4,4){\line(1,0){0.1}}
\put(16.7,4){\line(1,0){0.1}} \put(17,4){\line(1,0){0.1}}
\put(17.3,4){\line(1,0){0.1}} \put(16.1,1){\line(1,0){0.1}}
\put(16.4,1){\line(1,0){0.1}} \put(16.7,1){\line(1,0){0.1}}
\put(17,1){\line(1,0){0.1}} \put(17.3,1){\line(1,0){0.1}}

\put(18.6,4){\line(1,0){0.1}} \put(18.9,4){\line(1,0){0.1}}
\put(19.2,4){\line(1,0){0.1}} \put(19.5,4){\line(1,0){0.1}}
\put(19.8,4){\line(1,0){0.1}} \put(18.6,3){\line(1,0){0.1}}
\put(18.9,3){\line(1,0){0.1}} \put(19.2,3){\line(1,0){0.1}}
\put(19.5,3){\line(1,0){0.1}} \put(19.8,3){\line(1,0){0.1}}
\put(18.6,2){\line(1,0){0.1}} \put(18.9,2){\line(1,0){0.1}}
\put(19.2,2){\line(1,0){0.1}} \put(19.5,2){\line(1,0){0.1}}
\put(19.8,2){\line(1,0){0.1}}

\put(20.5,3.5){\vector(1,0){1}}

\put(22,2){\line(0,1){2}} \put(23,2){\line(0,1){2}}
\put(24.5,2){\line(0,1){2}} \put(25.5,2){\line(0,1){2}}
\put(22,2){\line(1,0){1}} \put(22,4){\line(1,0){1}}
\put(24.5,2){\line(1,0){1}} \put(24.5,4){\line(1,0){1}}
\put(22,3){\line(1,0){1}} \put(24.5,3){\line(1,0){1}}
\put(23.1,3){\line(1,0){0.1}} \put(23.4,3){\line(1,0){0.1}}
\put(23.7,3){\line(1,0){0.1}} \put(24.0,3){\line(1,0){0.1}}
\put(24.3,3){\line(1,0){0.1}} \put(23.1,2){\line(1,0){0.1}}
\put(23.4,2){\line(1,0){0.1}} \put(23.7,2){\line(1,0){0.1}}
\put(24.0,2){\line(1,0){0.1}} \put(24.3,2){\line(1,0){0.1}}
\put(23.1,4){\line(1,0){0.1}} \put(23.4,4){\line(1,0){0.1}}
\put(23.7,4){\line(1,0){0.1}} \put(24.0,4){\line(1,0){0.1}}
\put(24.3,4){\line(1,0){0.1}}
\end{picture}
\end{center}

Fig. \ref{suheis_FIGURE_magn5}  shows that that the magnetization
process of $S^z$, $T^z$ and $U^z$ are quite different. In phase
IV, the flippings from $|\overline \downarrow\rangle$ to other
three states possess positive  contribution to $S^z$ and $T^z$,
but negative contribution to $U^z$. In phase III, $S^z$ undergoes
polarization continually, while $U^z$ undergoes polarization but
$T^z$ undergoes anti-polarization. Since in the final phase, both
eigenvalues are zero and the spin magnetization are saturated,
this phase will not change for any further increase of the
external field.

The point of $g_t/g_s=0.5$,  $h=\ln 2$ is a three phase
coexistence point. It can be seen that there exist two phases
merely for $g_u/g_t<-1/2$, and phase IV transits into phase II
directly . Actually, the residual symmetry of
SU(2)$\times$U(1)$\times$SU(2) is restored when $g_t/g_s=-1$.

\newpage
\section{Brief summary}
\label{sec:discussion}

In above, we studied the magnetization properties of a SU(4)
spin-orbital chain in the presence of a generalized external field
that has three parameters due to the Cartan subalgebra of $su(4)$
Lie algebra has three generators. Those three parameters are
re-chosen so that to relate them with the spin $S$, orbital $T$
and their {\em product} $U$. We called them three Land\'e g
factors. Then all possible symmetry breaking and the corresponding
magnetization process induced by that external field are studied
respectively. The ground state phase diagram caused by  the
competition of quantum fluctuation and Zeeman-like effect are
studied by solving the Bethe-ansatz equations numerically. The
phase transition boundaries derived by studying the dress energy
equations analytically. The features of various phases and
transitions between them are explained in detailed in terms of
group theory analysis.

This work is supported by trans-century projects, Cheung Kong
projects of China Education Ministry, and NSFC No. 10225419, No.
90103022.

\newpage

\newpage

\newpage

\end{document}